\documentclass[3p,fleqn,authoryear,times]{elsarticle}
\usepackage{subcaption}
\usepackage{rotating}
\usepackage{tikz}
\usepackage{booktabs}
\usepackage{graphicx}
\usepackage{amssymb}
\usepackage{longtable}
\usepackage{amsthm}
\usepackage{multirow}
\usepackage{amsmath}
\usepackage[linesnumbered,ruled]{algorithm2e}

\usepackage{lineno}
\usepackage{array,calc}
\usepackage{lscape}
\usepackage{float}
\floatstyle{plaintop}
\restylefloat{table}
\usetikzlibrary{positioning}
\newlength\lena \newlength\lenb \newlength\lenc \newlength\lend
\newcolumntype{P}[1]{>{\centering\arraybackslash}p{#1}} 
\newcommand\mcii[1]{\multicolumn{10}{P{\lenb}|}{#1}}  
\newcommand\mciii[1]{\multicolumn{15}{P{\lenc}|}{#1}}

\usepackage{xcolor}
\usepackage{adjustbox}
\usepackage{subcaption}
\usepackage{tabularx}

\usepackage{array,calc}
\usepackage{makecell}

\newcommand\arash[1]{\textcolor{black}{#1}}

\usepackage{etoolbox}
\makeatletter
\patchcmd{\ps@pprintTitle}
  {Preprint submitted to}
  {Published in:}
  {}{}
\makeatother
\journal{Transportation Research Part C: Emerging Technologies}

\begin{document}

\begin{frontmatter}


\title{Decoding pedestrian and automated vehicle interactions using immersive virtual reality and interpretable deep learning}

\author[1]{Arash Kalatian\footnote{Corresponding author}}
\author[1]{Bilal Farooq}

\address[1]{Laboratory of Innovations in Transportation (LiTrans),

 Ryerson University, Toronto, Canada}

\begin{abstract}
To ensure pedestrian-friendly streets in the era of automated vehicles, reassessment of current policies, practices, design, rules and regulations of urban areas is of importance. This study investigates pedestrian crossing behaviour which, as an important element of urban dynamics,
 is expected to be affected by the presence of automated vehicles. For this purpose, an interpretable machine learning framework is proposed to explore factors affecting pedestrians' wait time before crossing mid-block crosswalks in the presence of automated vehicles.
To collect rich behavioural data, we developed a dynamic and immersive virtual reality experiment, with 180 participants from a heterogeneous population in 4 different locations in the Greater Toronto Area (GTA). 
Pedestrian wait time behaviour is then analyzed using a data-driven Cox Proportional Hazards (CPH) model, in which the linear combination of the covariates is replaced by a flexible non-linear deep neural network.
The proposed model achieved a 5\% improvement in goodness of fit, but more importantly, enabled us to incorporate a richer set of covariates.
A game theoretic based interpretability method is used to understand the contribution of different covariates to the time pedestrians wait before crossing.
Results show that the presence of automated vehicles on roads, wider lane widths, high density on roads, limited sight distance, and lack of walking habits are the main contributing factors to longer wait times. Our study suggested that, to move towards pedestrian-friendly urban areas, educational programs for children, enhanced safety measures for seniors, promotion of active modes of transportation, and revised traffic rules and regulations should be considered.
\end{abstract}


\end{frontmatter}

\section{Introduction}
\label{S:intro}
With a technological revolution in mobility on the way, unprecedented changes in urban areas are expected to happen. Before switching to Automated Vehicles (AVs), a careful and detailed investigation is required to analyze their impacts on future mobility patterns, travel behaviour and street design. As a part of human behaviour that is expected to change, interactions between different road users are of importance as they affect safety, efficiency, congestion, quality of service and public trust in transportation systems. In particular, the interaction that we seek to investigate in this study is that of pedestrians, as the most vulnerable road users, and vehicles. Recent instances of AV-pedestrian collisions, e.g. Uber's test AV fatal incident in Tempe, Arizona, and Navya SAS automated bus accident in Vienna, Austria, reveal the vital importance of such investigations. In an urban space dominated by rule-obeying automated vehicles that always stop for pedestrians even at mid-block unsignalized crosswalks, an emphasis on investigating this type of crossing is much needed~\citep{millard2018pedestrians}.
Several questions arise during this investigation: What factors, in terms of traffic parameters, rules and regulations, policies, design, and practices need to be taken into account before the transition towards automated urban environments? Would pedestrians behave differently confronting the yet fairly unknown phenomenon of automated vehicles? And how can we smooth this transition? Moreover, behaviour of pedestrians depend on various factors, e.g. their sociodemographic information, traffic parameters, environmental and lighting conditions, etc~\citep{rasouli2019autonomous}. How can we identify them and their contributions to pedestrian behaviour?

\begin{figure}[!h]
    \centering
    \includegraphics[scale=0.8]{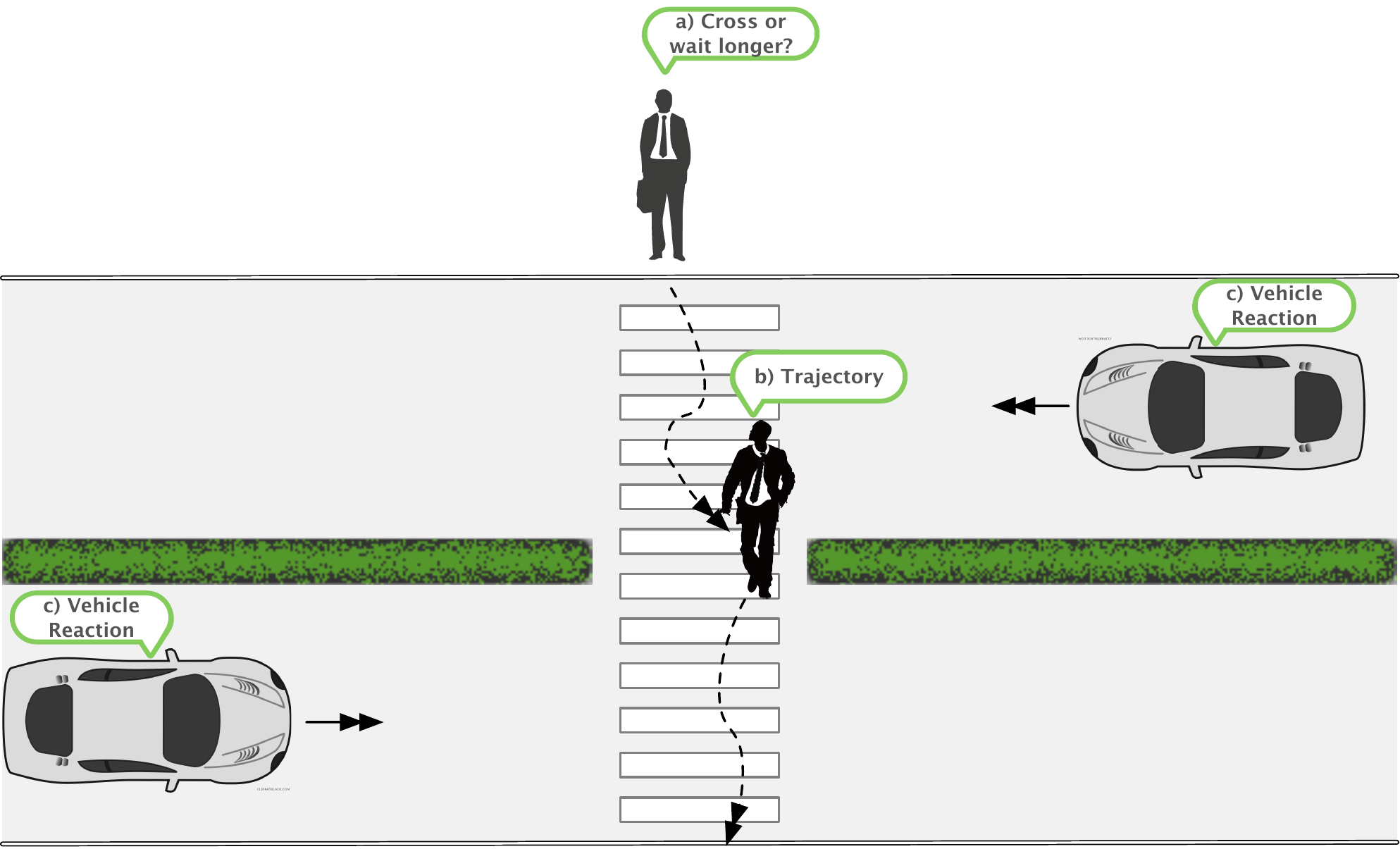}
    \caption{Schematic representation of vehicles and pedestrians interactions}
    \label{fig:interaction}
\end{figure}

Figure~\ref{fig:interaction} depicts a simplified schematic framework of behaviours of different agents involved in a mid-bloc unsignalized cross of a pedestrian. A pedestrian's behaviour in this context can be roughly abstracted into two parts: a) waiting on a sidewalk, during which a pedestrian intends to cross or decides to wait further until they feel more comfortable to cross, and b) crossing the street, which results in a pedestrian trajectory. The driver/vehicle reacts according to its observation of the environment and prediction of pedestrian's behaviour. In this study, we unfold the first part by analyzing pedestrians' waiting time and investigating factors affecting it under different conditions. By focusing on safe crosses of pedestrians, we try to understand what parameters, in term of pedestrians' sociodemographic information, rules and regulations, traffic measures, environmental and vision conditions, etc., affect the time it takes for the pedestrian to start crossing the street. We then propose and discuss possible practices towards transition to pedestrian-friendly automated environments considering the wait time of pedestrians. Wait time of pedestrians throughout this study is associated with safe crossings of pedestrians, without incurring any incidents or discomfort. Shorter wait time before a safe cross can also be associated with more trust in the vehicles, as opposed to longer wait times implying more hesitations to cross safely~\citep{jayaraman2019pedestrian}. Thus, we expect pedestrians who feel more comfortable and confident in the environment to have shorter waiting times.

Obtaining data for studies involving pedestrian crossing behaviour is often difficult, as it could be dangerous to participate in various crossing scenarios under different conditions. Related studies mainly rely on data extracted from video footage, which often fail to capture details required for policy suggestions and best practices. Such data are often collected passively from environments with either no or a limited form of control on variables of crossing scenarios. To tackle the difficulties of data collection for studies on futuristic subjects, and to elicit more realistic responses than stated preferences surveys, a virtual reality experiment is designed and implemented in this study, involving around 180 participants over a period of 5 months. 

To analyze the data, a neural network based survival model is developed, and the results are compared to traditional survival models. To enhance the performance of the model and add to its explainability, a feature selection method and a post-hoc game theoretic based interpretability method are added to the framework, respectively. Various researchers have studied the technology, adoption rate, demands, etc. of automated vehicles. However, according to a survey on recognizing the barriers to cities' AV efforts, the lack of clarity on issues that require city actions is pointed out as one of the biggest obstacles for preparing cities for automated vehicles~\citep{kinaSUR}. We try to address such issues by providing detailed insights from our experiments, models, and results.

This paper builds upon our previous work~\citep{kalatian2019deepwait} where an initial deep learning model was proposed for wait time. The current study has been extensively enhanced compared to the earlier work by adding a broader analysis of the data and data collection process, detailed explanations on the virtual reality experiments and their design, more consistent and systematic comparison of the models, a comprehensive analysis of deep learning models' interpretability, and policy and practical recommendations based on the results. We developed suggestions and practical implications which can be useful for urban planners, AV manufacturers, transportation researchers and policy makers. In short, contributions of the current study to the transportation research community can be summarized as follows:
\begin{itemize}
    \item Introducing, developing and utilizing a large-scale virtual reality experiment design and data collection procedure, which is of use for the experiments involving futuristic scenarios as well as situations that might be dangerous for participants to execute in real life.
    \item Developing a data-driven survival model, as well as its detailed and systematic interpretation to analyze the contributing factors to the wait time of pedestrians.
    \item Recommending policy suggestions and practical implications on both virtual reality data collection and pedestrian crossing behaviour. Future studies intending to use virtual reality for data collection can use suggestions to better manage their available resources, and policy makers can use our recommendations as either a validation tool or a ground for deciding on the issues of cities concerning pedestrians and/or automated vehicles.
\end{itemize} The rest of this paper is organized as follows: after a brief review of the literature on waiting time studies, pedestrian and automated vehicles interactions, neural network-based survival analysis and machine learning interpretability in section~\ref{S:lit}, the model is described in detail in section~\ref{S:meth}, along with the methodology implemented for feature selection and model interpretability. The data collection procedure, from experiment design to data summary is then presented in section~\ref{S:data}. Model results and analysis are then provided in section~\ref{S:results}, followed by a discussion of practical implications and policy insights in section~\ref{S:discuss}. Finally, a conclusion of the research and the future directions of the study will conclude this paper in section~\ref{S:Con}. 

\section{Background}
\label{S:lit}
First, a general overview of the research on pedestrian crossing behaviour is presented. As per the scope of the study, we mainly focus on studies on crossing behaviours at unsignalized and mid-block crossings. Survival models have primarily been used in such studies. Thus, a brief review on the background of survival analysis methods and their transformations using deep learning is provided. Then we discuss the literature on interpretability of machine learning models, especially those applicable to transportation research. As neural networks are generally considered a \textit{black box}, developing a systematic interpretability method is essential in our study to recognize contributing factors to pedestrian wait time and infer insights for practical implications of the results. Finally, relevant research studies on the use of virtual reality as a tool for data collection are discussed.
\subsection{Pedestrian unsignalized crossing behaviour}
Studies on gap time acceptance form a large part of pedestrian crossing behaviour research studies at unsignalized intersections and mid-block crossings. \cite{das2005walk} found that people waiting on medians accept shorter gaps more easily, compared to those waiting on the curb sides. Distribution of gaps was estimated in this study using parametric and non-parametric approaches. \cite{oxley2005crossing} focused on the effect of age on the ability to choose safe gap time. First, ANOVA was used to compare wait times of participants. According to their results, the primary contributing factor in deciding to cross, which can be interpreted as waiting time, appeared to be the distance to upcoming vehicle and time of arrival. Moreover, elderly participants appeared to select more unsafe gap times, given that their walking speeds were slower. In a relatively more recent work, \cite{kadali2014evaluation} used artificial neural networks to estimate pedestrians' gap acceptance at mid-block unmarked crossings and compared its performance to multiple regression models. Their results showed that ANN has a better prediction performance, being able to consider the effects of more number of variables. However, the authors mentioned the strength of regression models in such cases due to their ability to reflect the significance of contributing variable.

Systematic modelling of pedestrian wait time using cox proportional hazards (CPH) models dates back to early 2000's, where Hamed developed a CPH model to analyze the time pedestrians spend on the sidewalk before initiating a cross on mid-block crosswalks. Data used in this study were collected by observing and manual recordings, and interviews at pedestrian crossings. Among the factors analyzed, previous accident experiences, group size of people crossing, car ownership, trip purpose, gap time between cars and finally, age and gender were appeared to be the contributing factors to wait time of pedestrians in their sample size~\citep{hamed2001analysis}. Interestingly, parameters related to traffic were not found to be statistically significant in Hamed's study. In another application of survival models in pedestrian wait time analysis, \cite{wang2011individual} investigated the effects of personal characteristics on crossing behaviour using a parametric survival analysis. The effect of age, trip purpose, safety awareness and conformity behaviour appeared to be the contributing factors to pedestrian wait time in this study. 

\subsection{Interaction of pedestrians and automated vehicles}
Despite the long history of research activities on pedestrian crossing behaviour, studies on their behaviour in the presence of automated vehicles are yet relatively new. Analyzing communication techniques between the two involved agents form a large part of these studies~\citep{millard2018pedestrians,clamann2017evaluation,rasouli2017agreeing,mahadevan2018communicating}. Although the effect of communication systems among pedestrians and automated vehicles have appeared to be positive according to most studies, parameters that were shown to be important in traditional studies on human-driven vehicles are yet contributing factors to the behaviour of pedestrians in the context of automated vehicles~\citep{pillai2017virtual,clamann2017evaluation}.

Perceived risk of automated vehicles by the pedestrians has been another direction in the literature of pedestrians and automated vehicles. Generally, it is expected that different people would react and behave differently to the concept of automated vehicles. By conducting several questionnaire surveys, \cite{deb2017pedestrians} found younger people, males, people from urban areas, and people with higher personal innovativeness to have higher acceptance towards AVs. Similar results on the effect of age, gender, and risk-taking personality were achieved in questionnaires surveyed by \cite{hulse2018perceptions}. \cite{jayaraman2019pedestrian} investigated trust in automated vehicles by conducting a virtual reality study. By studying the role of traffic signals and driving behaviour on the trust of AV by pedestrians, they found that trust in AVs is strongly correlated with pedestrian gaze, pedestrian distance to collision, and pedestrian jaywalking time. Reading through the relatively new literature on pedestrian behaviour in the presence of AVs, a major research gap seems to exist, which we try to address in this study: Lack of research on the performance of variables that were significant in traditional crossing behaviour studies, within the new automated context. Variables such as street width, road design, and traffic variables have not been a major part of the research on pedestrian-AV interaction.  Failure to account for these variables before the transition towards automated environments can lead to inefficient or dangerous designs, where vehicles regard pedestrians more conservative or reckless than they really are~\citep{rasouli2019autonomous}.

\subsection{Survival models and data-driven machine learning}
In survival analysis, the aim is to analyze the time until an event occurs. A collection of statistical modelling methods can be designed for this purpose. The most common methods for survival analysis are the Kaplan-Meier model~\citep{kaplan1958nonparametric} and Cox Proportional Hazards (CPH) model~\citep{cox1972regression}.
The Kaplan-Meier model is a non-parametric model for estimating the survival function in homogeneous groups. This model is very easy to implement, but is unable to account for individuals. To take into account covariate vectors and compute survival functions for individuals, CPH model is a conventional solution. CPH model is a semi-parametric model that assumes the time component and the covariate component of hazard function to be proportional. Hazard function is defined as the risk of an event with explanatory variables (covariates) $Z$ to occur, at time $t$. Despite being the common method of survival analysis for years, the assumptions made in Cox Proportional Hazards are not always true and have limitations. It assumes a) a linear combination of covariates within the hazard function and b) a constant relative effect of covariates over time.
To address these issues, \cite{yu2011learning} introduced a Multi-Task Logistic Regression model. Their model can be interpreted as having different time intervals with different logistic regression models that estimate the probability of occurrence of an event in a interval. Although Multi-Task Logistic Regression models address some of the problems of the previous methods, they still remain linear, thus they cannot capture the nonlinear complexities within the data. Several researchers, particularly in the field of medical sciences, have tried to address this issue by implementing a deep learning approach \citep{faraggi1995neural,katzman2016deep,luck2017deep}.
\cite{faraggi1995neural} first incorporated a feed-forward neural network as a shift to nonlinear proportional hazards model. In their model, they replaced the linear combination of covariates with the output of a neural network with one output node. However, research followed by suggested that their network did not perform better than linear Cox Proportional Hazards models~\citep{mariani1997prognostic}. As the mentioned work were done prior to the outburst of modern deep learning algorithms, \cite{katzman2016deep} decided to test the performance of more recent deep networks on survival models. They added fully connected and dropout layers and outperformed the performance of previous linear cox proportional hazards models. As a suggested future direction, the authors suggested implementing architectures like Convolutional Neural Networks to enable estimating waiting time directly from medical images. 

Strong performances of survival analysis models in different areas, along with their relatively widespread usage in pedestrian wait time modelling, made them a suitable base model for our study. In an earlier study in our lab, we started analyzing crossing behaviour by studying distracted pedestrian's waiting time using Cox Proportional Hazards models, and we addressed various contributing distraction factors~\citep{kalatian2018cox}. Later, in~\citep{kalatian2019deepwait}, we introduced \textit{Deepwait}, which is a neural network based extension of CPH, empowered with a feature selection algorithm. However, despite the improvements in accuracy we obtained by using Deepwait, lack of a interpretability mechanism makes the application of such models limited, especially as learning the contributing factors are of vital importance. To better understand the problem of interpretability in machine learning models, and to be able to propose policy recommendations based on the results of our model, a review on the existing literature of machine learning interpretability is presented in the following subsection.
\subsection{Interpretation in machine learning}
Widespread application of machine learning methods for decision making and policy planning requires a high level of interpretability for often complex, dense, and deep networks. Despite high prediction accuracies obtained by machine learning models in recent years, interpretability and explainability of the algorithms are yet considered as a cumbersome and in some cases, even unnecessary task. Major arguments against interpretability of machine learning models are the notion that performance of a model is more important than model interpretability and, the burden of understanding the model can prevent adoption of machine learning models~\citep{lipton2016mythos}. Nevertheless, the trade off between model complexity and explainability may weigh in favor of using traditional explainable models in some areas where gaining information on \textit{why?} and \textit{how?} the model works is as important as the predictivity of the model.

To fill the existing void in machine learning models, various approaches to make machine learning models explainable have emerged in recent years. The \textit{black-box} nature of machine learning algorithms can be addressed by model-agnostic methods, which are post-hoc interpretability methods that rely solely on input and output of the models, disregarding their structure ~\citep{molnar2019interpretable}. Mainly practiced model-agnostic methods include sensitivity analysis plots~\citep{goldstein2015peeking}, gradient boosting feature importance methods~\citep{friedman2001greedy}, feature permutation methods~\citep{fisher2018all}, surrogate local interpretable estimators~\citep{ribeiro2016should}, and game theoretic based approaches~\citep{vstrumbelj2014explaining,lundberg2017unified}. In transportation, not so many studies have addressed the interpretability of the models despite the relatively large number of works on machine learning and deep learning applications in the field. In the interpretability of choice analysis for instance, \cite{hagenauer2017comparative} used a permutation-based variable importance method upon their DNN model to analyze the performance of the model when the values of a variable changes randomly. In \citep{wang2018deep}, by considering the inputs to the Softmax layer as the utility functions, authors compute economic information using DNNs. They conclude that economical information aggregated over the population can be reliably derived from DNN. However, disaggregate information for individuals still lacks the reliability acquired by discrete choice models. \cite{wong2020bi} used Jacobian determinant of generative models to calculate the elasticity of mode choice with respect to different explanatory variables. Jacobian matrix, in this study, was generated for each instance of all the conditional outputs, and density of elasticises was estimated across the data points. Use of Jacobian matrix, however, is computationally expensive and not guaranteed to observe the interactive effects of variables. For prediction and estimation tasks, \cite{das2019interpretable} used partial dependency plots to intuitively show the effects of different variables and detect contributing factors on predicting traffic volumes. Most of the existing work in transportation rely on intuitive and illustration-based interpretability frameworks without considering the interactive effects of variables in a systematic manner. To address this issue, more recent and advanced methods of game-theoretic based interpretability methods will be used in this study to find the contributing factors to pedestrians' waiting time before crossing.
\subsection{Virtual reality as a tool for data collection}
To study the perception of people, pictures, maps and videos have been used in the past. Recent developments in Virtual Reality (VR) environments have made investigation of human perception and behaviours in controlled conditions possible. Combination of the physical environment and virtual elements, i.e. information or images, broadens the opportunities for content delivery~\citep{farooq2018virtual,jennett2008measuring,animesh2011odyssey,nah2011enhancing,faiola2013correlating}.

One of the main concerns that are raised for using virtual reality is the level of realism involved and how the results obtained through VR resemble those obtained from behaviours in real life. Several researchers have investigated such comparison. In studies on pedestrian behaviour, \cite{bhagavathula2018reality} compared different elements of pedestrian behaviour in virtual reality and real environments. Their comparison resulted that crossing intention, perception of safety, perception of risk and perception of distance were not significantly different in the two environments. On the other hand, it appeared that perception of speed in the two approaches of data collection were different among the participants. Similarly, \cite{deb2017efficacy} compared the behaviour of pedestrians crossing signalized intersection and concluded that both objective and subjective measures were similar in the participants in the two environments. However, an 11\% failure in completing the VR experiments was observed during their data collection, due to motion sickness. In another study focused on body movements of pedestrians crossing in virtual reality environments, \cite{kalantarov2018pedestrians} concluded that wait time measures in their virtual realty experiment was in line with the laboratory studies and field observations.

Regarding pedestrian behaviour in the presence of automated vehicles, \cite{jayaraman2019pedestrian} analyzed trust in AVs using virtual reality data. Related to the aim of our study, they hypothesised that trust in AV will decrease pedestrian wait time. However, their modeling results showed the opposite. Studies using VR often lack a large number of participants. Thus, the conclusion may be biased towards the limited number of people in the experiment. In another study using VR experiments, \cite{pillai2017virtual} found weather conditions and cultural differences affecting pedestrians' perception of automated vehicles. In this study, by running a relatively large data collection campaign, and by focusing on a specific type of crossing, we try to fill this gap and verify the applicability of virtual reality for pedestrian research data collection in larger scales.
\section{Methodology}
\label{S:meth}
Pedestrian wait time before they start crossing can be modelled as the time it takes before an event occurs. Thus, survival analysis is a popular and powerful tool used for this purpose. In the following, first traditional survival models, and in particular, Cox Proportional Hazards (CPH) models are explained. We used CPH as the base model for comparison in this study. Then, the data-driven deep neural network version of CPH is proposed as the framework used in this study, which enables incorporating more covariates while increasing goodness of fit. Finally, interpretability mechanism used to explain model results and obtain practical implications is outlined.
\subsection{Survival analysis}
Survival analysis techniques are used in cases the time to occurrence of an event is of interest. Although originally developed for applications in medical science where the event is treatment of a disease or the death of the patients, survival methods have found their way in other fields, including transportation. By defining \textit{survival function} $S(t)$ as the probability that the event occurs after time $t$, and \textit{hazard function} $h(t)$ as the probability of the event occurring at time $t$, the following relationship can be established between the two:
\begin{linenomath}
\begin{equation}
\label{eq:1}
    S(t)=\exp({-\displaystyle \int_{0}^{t} h(z) dz)}
\end{equation}
\end{linenomath}
Different methods have been proposed for estimating hazard function, parametric, semi-parametric and non-parametric. Among them, Cox Proportional Hazards (CPH) model is probably the most widely used one, which enables considering the effects of covariates on the hazard function~\citep{cox1972regression}. Equation \ref{eq:2} shows the hazard function formula in CPH.
\begin{linenomath}
\begin{equation}
\label{eq:2}
    h(t|Z)=h_0(t) \times \exp({\displaystyle \sum_{i} \beta_i Z_i})
\end{equation}
\end{linenomath}
In the above equation, $h_0(t)$ is the baseline hazard, which is the hazard function regardless of the values of the covariates and is a function of time, and $\beta_i$ and $Z_i$ are the coefficient and value of covariate $i$, respectively~\citep{therneau2013modeling}. In the terminology of survival analysis, ${e^{\textstyle \sum_{i} \beta_i Z_i}}$ is called partial hazard function or risk function, which is representative of the effects of values of covariates on the hazard function and is independent of time. As it can be seen in equation ~\ref{eq:2}, the log-partial hazard function in CPH is in a linear combination of covariates. To find the coefficients in CPH, the partial likelihood shown in equation~\ref{eq:3} is maximized.
\begin{linenomath}
\begin{equation}
\label{eq:3}
\mathcal{L}(\beta) = \prod_{k\in instances}{\exp({\displaystyle \sum_{i} \beta_i Z_{ik}})} / {\displaystyle \sum_{j: T_j > T_k} \exp({\displaystyle  \sum_{i} \beta_i Z_{ij}})}
\end{equation}
\end{linenomath}
Where, index $j$ in the denominator selects the instances that are still at risk when event of instance $k$ occurs. In this study, we use CPH with linear log-partial hazard function as the base model, in which an event is defined as initiation of a cross by a participant. The simplistic assumption of linearity of log-partial hazard function is what drives us to improve and update the model to account for today's complex data and advanced modelling tools available.  
\subsection{Data-driven survival analysis}
\begin{figure}[!h]
    \centering
    \includegraphics[scale=0.95]{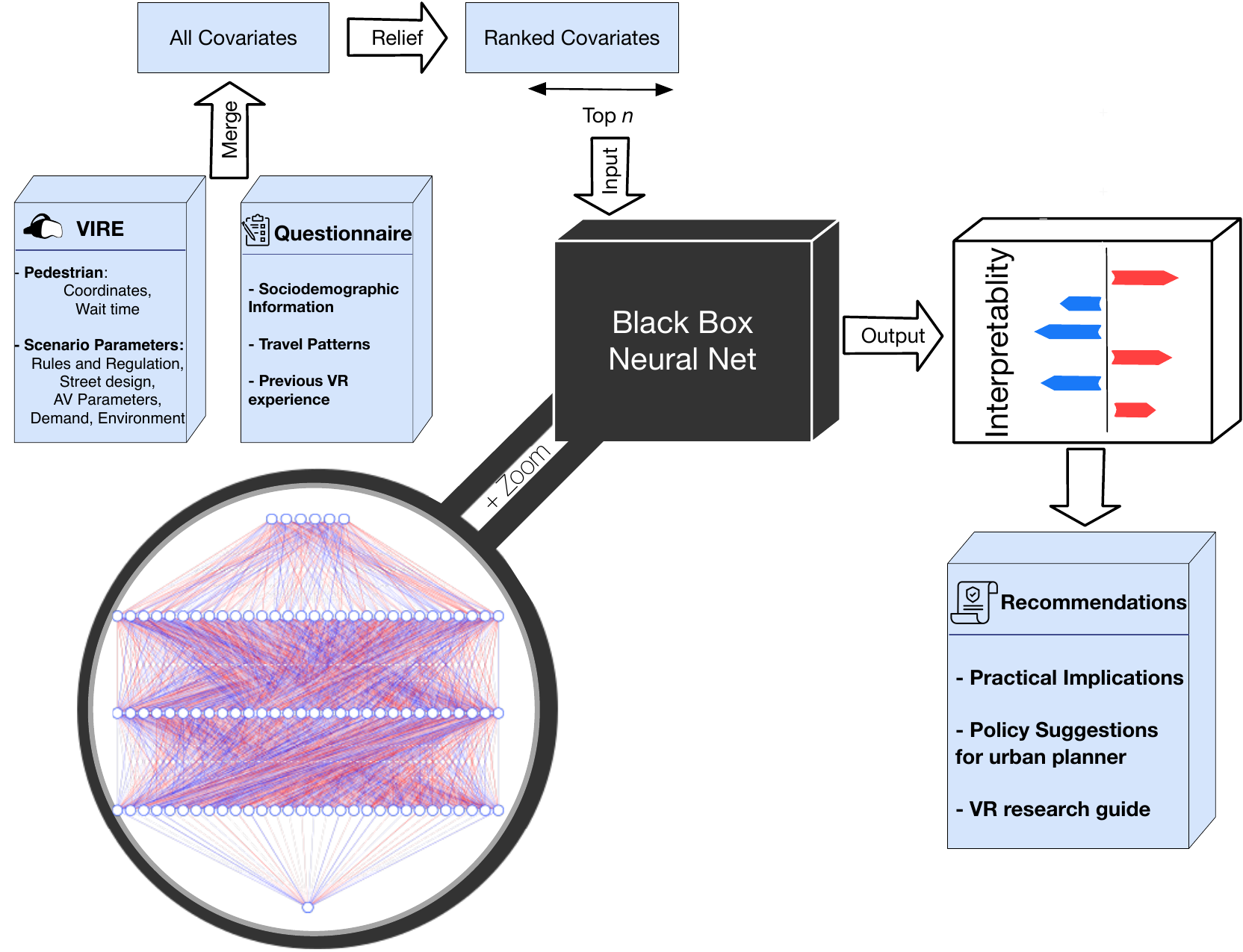}
    \caption{Framework of the study}
    \label{fig:framework}
\end{figure}

Rapid and ubiquitous emergence of data-driven approaches in recent years provides an opportunity to further improve the models for survival analysis. Due to the availability of complex and high-dimensional data today, traditional methods may not be capable of capturing nonlinearities within data. Thus, we build upon the classic CPH (Equation \ref{eq:2}) and replace the linear log-partial hazard function ($\ \sum_{i}\beta_i Z_i$) with a deep neural network, $g(w)$ \citep{kalatian2019deepwait}. The proposed framework used in this study is outlined in Figure~\ref{fig:framework}, and the model components are explained in the next two subsections.

\subsubsection{Feature importance ranking}
High-dimensionality of modern data sources has led researchers and machine learning engineers to incorporate feature selection methods to prevent overfitting and performance degradation~\citep{alelyani2018feature}.  Although deep neural networks are known for their capability of detecting correlation among features, a Variance Inflation Factor (VIF) analysis and feature selection algorithm is implemented in the preprocessing step to 1. develop a base case cox proportional hazards model, and 2. compare the effectiveness of deep survival models with and without data preprocessing.   

Initially proposed by \cite{kononenko1997overcoming} for classification, Relief family of algorithms consider the ability of variables to distinguish among instances that are close to each other,and rank their quality based upon this ability. RRELIEFF~\citep{robnik1997adaptation}, which is an adaptation of Relief for using in regression, is used in this study to prioritize covariates, and reduce the dimensionality of the dataset collected. Based on Relief in its simplest form, the importance of a covariate in a classification task increases, if its value is different for instances from different classes, and decreases if its value is different for instances of the same class. RreliefF, which is the regression version of the family, replaces the classes with a probability defined based on the relative distance between the target values of the instances. In mathematical terms, importance weight of covariate $f$, $W_f$, is calculated based on the following equation~\citep{robnik2003theoretical}:
\begin{linenomath}
\begin{equation}
\label{relief}
    W_f = ({P_{v|f}\times P_f})/{P_v}  - ({(1-P_{v|f})\times P_f})/({1-P_v})
\end{equation}
\end{linenomath}
Where:
\begin{itemize}
    \item $P_f$ = P(different value of $f$ in nearest instances)
    \item $P_v$ = P(different target value in nearest instances)
\end{itemize}
\subsubsection{Hazard function}
The formulation of modified hazard function in our proposed model is shown in equation~\ref{eq:5}:
\begin{linenomath}
\begin{equation}
\label{eq:5}
    h(t|Z)=h_0(t) \times \exp({g_w(Z)})
\end{equation}
\end{linenomath}
where $g_w$ is the dense neural network with weights $w$. Top $n$ covariates ($Z_n$), based on the feature importance ranking, will be imported into the network's input layer.  After a number of fully connected hidden layers, each followed by dropout layers and batch normalization to prevent overfitting to the training data, the output of the network $g(Z_n)$ will be calculated as the value of the last one-node layer, and will be used as our estimation of log-partial hazard. 

In the deep network described, network architecture parameters i.e. number of covariates to be selected, number of hidden layers, number of nodes in each layer and dropout rate as well as optimization parameters, including learning rate decay and momentum, are used as the hyperparameters to be found. The loss function to be minimized in network training is derived from the partial likelihood function of CPH. As shown in equation~\ref{eq:6}, network's loss function would be the average negative logarithm of CPH's partial likelihood:
\begin{linenomath}
\begin{equation}
\label{eq:6}
L_w=-({1}/{N})\times \displaystyle \sum_{k\in instances} \left(g_w(Z_{nk}) - \log\displaystyle  \sum_{j: T_j > T_k} \exp({g_w(Z_{nj})}) \right)     
\end{equation}
\end{linenomath}
where $N$ is the total number of events or instances, and $Z_{nk}$ is the is the value of top $n$ covariates for instance $k$.

\subsection{Model interpretability}
Transition from regression models to neural-based models comes with the cost of losing useful information pertaining to the effects of covariates on the model output. Within traditional CPH models, the effect of each covariate on the log-partial hazard function can be estimated by looking into covariate coefficients. A greater coefficient for a covariate means an increase in log-partial hazard function, and subsequently in hazard function. In other words, a positive coefficient for a covariate means that greater value of that covariate results in higher probability of the event to occur at any time, thus a shorter wait time. By using z-scores and p-values, the significance of variables can be measured in traditional based CPH models. The case is different for neural networks, however, as it is impossible to track the nonlinearities involving multiple variables at the same time and in different layers. 

To address this issue, SHAP (SHapley Additive exPlanations) is used in this study. Introduced in \citep{lundberg2017unified}, SHAP is a game theoretic model explanation method based on Shapley Values. In game theory, Shapley value is a method to \textit{fairly} distribute the payoff of a job done by a coalition of players with different skills, to the players. By replacing the payoff with model prediction and players with features, SHAP calculates the average marginal contribution of each feature to model prediction for each instance. In mathematical terms, Shapley value of each feature $i$ is calculated by equation~\ref{eq:7} as follows~\citep{lipovetsky2001analysis}:
\begin{linenomath}
\begin{equation}
\label{eq:7}
\Phi_i = \sum_{S \in F \setminus \{i\}} \frac{|S|! (|F| - |S| -1)!}{|F|!} \left(
g_{S\cup\{i\}}(Z_{S\cup\{i\}}) - g_S(Z_S) \right)
\end{equation}
\end{linenomath}
In which $S$ is a subset of all features $F$, $g_{S\cup\{i\}}$ is the model trained using a subset with feature $\{i\}$ present, and $g_S$ is the model trained with the feature $i$ missing. One of the advantages of using Shapley values over other interpretation methods is that it considers the fact that the contribution of a feature depends on the values of other features, thus the contributions are computed for all possible subsets. In their paper, \citet{lundberg2017unified} show that SHAP solution satisfies required conditions to fairly distribute the payoff to players. By using Shapley values, importance of the features can be estimated as the average absolute Shapley values of features over the whole instances. 
\section{Controlled experiment and data}
\label{S:data}
\subsection{VIRE}
Virtual Immersive Reality Experiment (VIRE) is used in this study as the tool to develop the controlled experiments and collect the required data. We selected virtual reality, as our experiments involve situations that are either dangerous in reality and may result in disastrous outcomes, or are futuristic scenarios that are not yet feasible to implement. On the other hand, stated preferences surveys, which are the conventional tool of data collection in similar studies, cannot provide naturalistic datasets as participants have no prior exposure to automated vehicles.

 VIRE is an in-house developed virtual reality framework designed for experiments in controlled environments~\citep{farooq2018virtual}. The unique feature of VIRE is that the virtual objects are designed such that they react to the participant's actions. For instance, if a participant is crossing the road, based on their location, the approaching vehicle may slow down or completely stop, thus giving a dynamic, immersive, and realistic experience. A scenario is generated based on a set of variables and then it is projected onto the eyes of participants through an immersive head mounted display. Participants, finding themselves on a simulated 3D sidewalk, are asked to start crossing when they feel it is comfortable to do so. During the experiments, data on the movements of pedestrians are recorded using motion sensors and the reactions of virtual objects are computed in real-time.

Each scenario is defined by 9 controlled variables, which are selected based on related literature on future urban streets~\citep{kinaSUR}, while considering the available facilities. These variables are categorized as follows: 
\begin{itemize} 
\item \textbf{Rules and regulations:} speed limit, minimum allowed gap time between vehicles 
\item \textbf{Street design:} lane width, type of road 
\item \textbf{Automated vehicles:} traffic automation status, number of braking levels
\item \textbf{Demand:} traffic flow (arrival rate)
\item \textbf{Environmental conditions:} time of day, weather.
\end{itemize}
Multiple levels are defined for each of the controlled variables, as presented in Table~\ref{tab:levels}. The standards of speed limit, suggested minimum gap between cars and lane width at the time of experiment in a typical road in downtown Toronto were 50 \textit{km/hr}, 2 \textit{sec}, and 3 \textit{m}, respectively. To take into account future possible modifications to the current standards, one of the levels for each of these three variables are set in our experiment to have the same value of a typical Toronto street standard, with the other two levels considering possible changes. Number of braking levels, which is only defined for automated vehicles, represents the level of smoothness of braking system of the AVs. Having a greater number of levels shows a more smooth deceleration when AVs face a barrier. Level 1 system for braking means the vehicle decelerated with a constant rate as long as a barrier (crossing pedestrian in our case) is observed on the road. In level 2 systems for braking, the AV first decelerates to a speed equal to the half of its initial speed in half of their initial distance to the barrier. If the barrier still exists, the AV will change the deceleration to fully stop for the barrier in the rest of the distance. Similar pattern exists for level 3 braking system with three deceleration rates calculated and applied. On the other hand, braking for all human-driven vehicles are defined as level 1. \arash{Details on how the braking levels work, as well as sample velocity profiles of a hypothetical vehicle under different braking levels are provided in \ref{A:brake}.} Simulated traffic on the road can either be fully human-driven, fully automated or a mixture of automated vehicles and human-driven cars. In the experiments, users distinguish human-driven and automated vehicles by observing the absence of a driver in the driving seat in AVs, and a box-shaped LiDAR sensor on top of them. Similar to \citep{jayaraman2019pedestrian}, participants get familiar with the shape of AVs in the training before the experiments, so that they can tell the difference between the two types of vehicles. Moreover, AVs and human-driven cars are different in braking systems, which might not be realized by the participants, unless they do the experiment as such that a vehicle has to brake for them. As for the demand on the road, values are extracted from flow rates on a typical road in downtown Toronto during off hour and peak hour hourly average (530 and 1100 \textit{veh/hr/lane} respectively) and the mean value of these flow rates (750 \textit{veh/hr/lane}). Two flow variables are also derived based on the controlled variables and used in the analysis: distance between cars and traffic density. Finally, the environmental conditions in the simulation are designed as such that participants experience day and night time of the day, as well as clear and snowy weather, in which the sight distance of pedestrians is affected. All the collected variables using VR, as well as those collected through the questionnaires before the experiments are described in \ref{A:var} While doing the experiments, coordinates and head orientations of participants are recorded in intervals of 100 milliseconds, as well as the coordinates of simulated vehicles. 

\begin{table}[!h]
\footnotesize
\caption{Controlled variables and their levels}
\setlength\extrarowheight{2pt}
\centering
\begin{tabular}{|l|l|*{30}{p{\lena}|}}
\hline
\bf{Factor}&\bf{Variable}           & \multicolumn{30}{c|}{\bf{Levels}}\\
\hline\hline  
\multirow{2}{*}{\textbf{\begin{tabular}[c]{@{}l@{}}Rules and\\ regulations\end{tabular}}}&Speed limit (km/h) & \mcii{30} & \mcii{40} & \mcii{50} \\
\cline{2-5}
&Minimum allowed gap time (s) & \mcii{1} & \mcii{1.5} & \mcii{2} \\  
\hline
\multirow{2}{*}{\textbf{Street design}}&Lane width (m) & \mcii{2.5} & \mcii{2.75} & \mcii{3} \\  
\cline{2-5}
&Road type & \mcii{1-way} & \mcii{2-way} & \mcii{2-way with median} \\  
\hline
\multirow{2}{*}{\textbf{\begin{tabular}[c]{@{}l@{}}Automated\\ vehicles\end{tabular}}} &No. of braking levels &  \mcii{1} & \mcii{2} & \mcii{3} \\  
\cline{2-5}
&Traffic automation status& \mcii{Fully human-driven} & \mcii{Mixed traffic} & \mcii{Fully automated} \\  
\hline
\textbf{Demand}&Arrival rate (veh/hr) & \mcii{530} & \mcii{750} & \mcii{1100} \\  
\hline
\multirow{2}{*}{\textbf{Environmental}}&Time of day & \mciii{Day} & \mciii{Night}  \\
\cline{2-5}&Weather & \mciii{Clear} & \mciii{Snowy}  \\
\hline
\end{tabular}
\label{tab:levels}
\end{table}

Scenarios are then generated based on the above-mentioned controlled variables. Each possible combination of different levels of controlled variables is a potential scenario. To select the scenarios for the experiments, a design of experiment is conducted, which is described in detail in the next section. 

\subsection{Design of experiment}
To study the effects of various different attributes on pedestrians' crossing behaviour, a design of experiment (DoE) needs to be conducted. The first and simplest method for DoE is full factorial design, in which after assigning different levels to each attribute, all possible combinations of different attributes' levels are considered for the experiment. In our case, however, this is an impossible approach to take as there would be a total of 8,747 possible combinations based on the 9 attributes we defined and their levels. To overcome this problem, several \textit{Fractional Factorial Designs} have been developed based on the objectives of the design, models to be developed and the amount of prior information available~\citep{cavalcante2011bayesian}. In fractional factorial designs, a subset of choice tasks are selected. The most widely used strategy for selecting a subset of tasks is orthogonal design, in which tasks are selected in a way to produce zero correlation between attributes~\citep{ortuzar1994modelling}. However, orthogonality may limit the experiment design by putting constraints in terms of the number of runs required and possible settings for factor levels. Recent studies on the topic have led to \textit{Optimal Designs}, which focus on the efficiency of the experiment, rather than orthogonality~\citep{rose2009constructing}. In optimal and efficient designs researchers try to find the efficient design in terms of a selected measure of quality for parameters' estimates. In the most widely used optimal design, D-Optimal design, for instance, variances and covariances of parameter estimates determine attribute level combinations~\citep{atkinson2007optimum}. For the purpose of data collection for this study, we use an optimal design of experiment, as it allows experiments to be conducted in a more flexible manner in terms of number of runs required and tasks to be observed. 

In D-Optimal designs, a number of assumptions and parameter estimates for the model are required in order to calculate covariance matrix, as this matrix varies based on the model used. In order to deal with parameter priors required in these methods, two approaches can be taken. First approach is to consider parameter priors to be zero, which is called the null hypothesis~\citep{street2005quick}. This assumption indirectly results in the orthogonality of null hypothesis D-Optimal designs. In addition, it is assumed in this type of methods that the model used is Multinomial Logit (MNL). Not requiring a priori knowledge on parameter estimates makes this method useful when no prior information is available. The second approach, on the other hand, assumes that some estimates are available on the values of parameters. Rose and Bliemer suggested that a priori parameter estimates can be obtained from: the literature, pilot studies, focus groups and expert judgments~\citep{rose2008designing}. The main advantage of this approach is that it enable researchers to use any model type, in contrary to the former approach that was applicable only to MNL.

In this study, we obtained estimates on a priori parameters based on a trial experiment on a limited number of participants. To do so, a demo of the data collection was conducted with 5 participants, each going through 30 randomly selected scenarios to get an initial idea on the model parameters. As the objective of the data collection was to obtain required data to model pedestrian wait time, we assumed a cox proportional hazards model (CPH) for waiting time~\footnote{CPH models and the reason we selected them for our modelling are discussed in details in section~\ref{S:meth}.}. Thus, a design of experiment was developed based on the CPH model on the trial data, and top scenarios based on importance weights were selected for our experiment. In total, 90 different scenarios were selected. The design of experiment formulation for wait time modelling, which is the topic of interest in this study is presented in~\ref{A:design}

\subsection{Data collection campaign}
\begin{figure}[!ht]
\begin{adjustbox}{varwidth=\textwidth,fbox,center}
\centering
\begin{subfigure}{0.8\linewidth}

    \begin{subfigure}[b]{0.5\textwidth}
         \centering
         \includegraphics[trim=100 100 100 175,clip,
         scale=0.06]{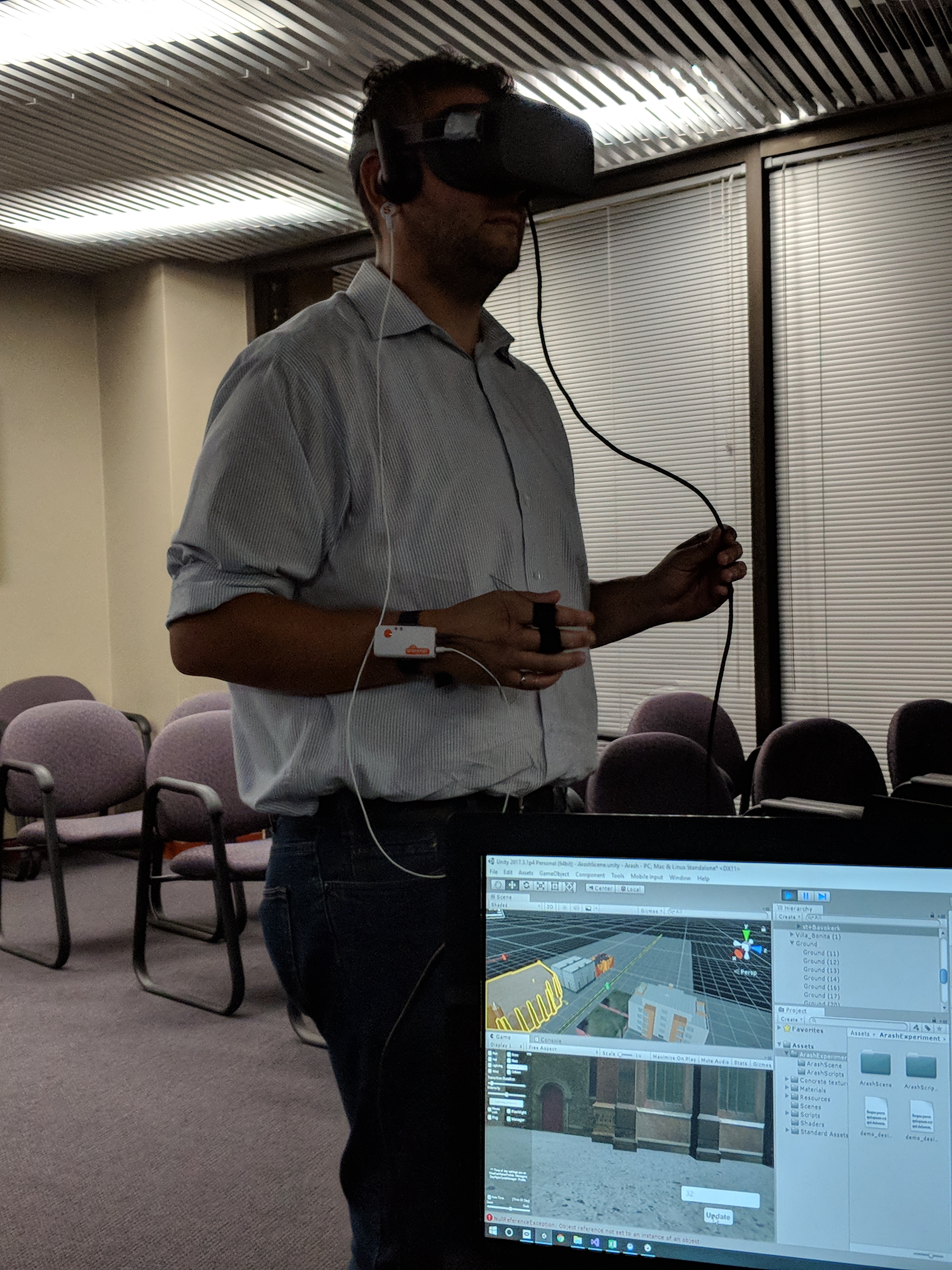}
         \caption{A participant in Toronto City Hall}
     \end{subfigure}
     \hfill
     \begin{subfigure}[b]{0.5\textwidth}
         \centering
         \includegraphics[scale=0.4]{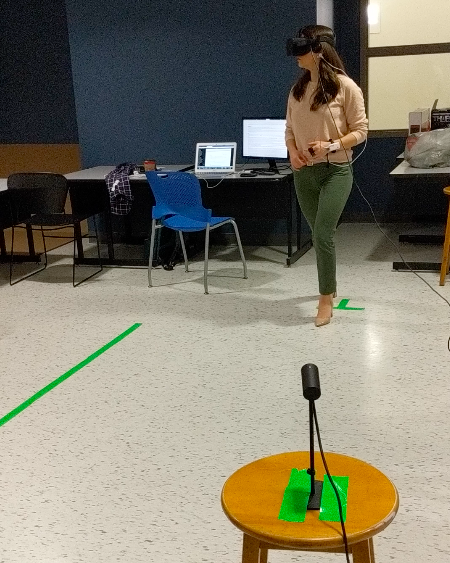}
         \caption{A participant at Ryerson University}
     \end{subfigure}
\end{subfigure}\\[2ex]
\begin{subfigure}{0.8\linewidth}

\centering
\includegraphics[scale=0.1]{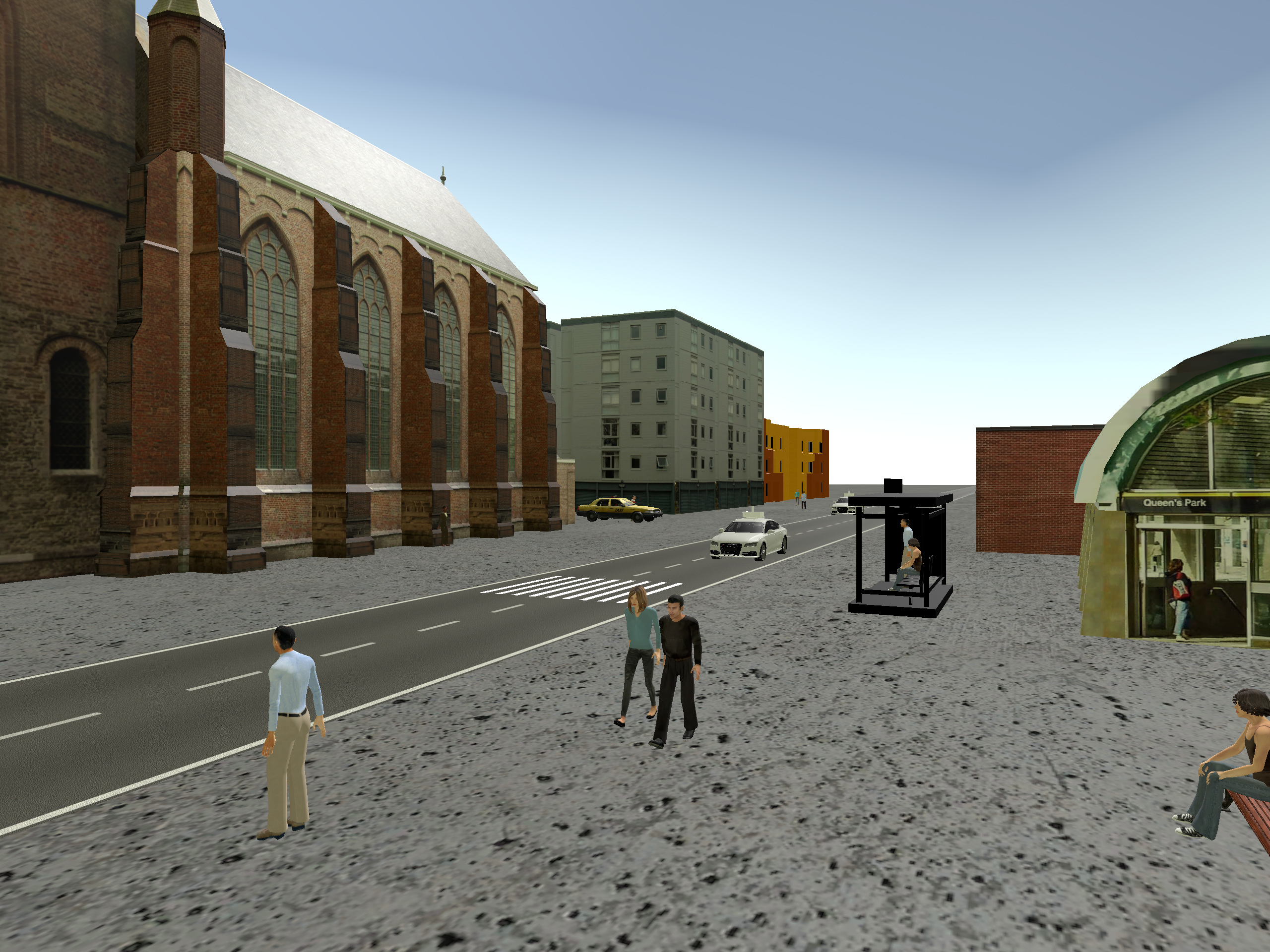}
\caption{A virtual sample environment}
\end{subfigure}
\end{adjustbox}

\caption{A sample view of experiments}
\label{fig:vire}
\end{figure}

Data collection campaign for this study started in April 2018, as a trial experiment with a limited number of participants, and then continued with main experiments until September 2018. In total, 180 people participated in the experiments, 160 of which succeeded to complete at least one of their tasks in the experiment. These 160 participants consisted of 113 adults, and 47 kids and teenagers. To make the data as inclusive and heterogeneous as possible, we conducted our experiment in four different locations. Started at Ryerson University with post-secondary students, we then moved to Toronto City Hall and North York Civic Center to include professionals in the field who were more familiar with the nature of the experiments and the questions we were trying to answer. We repeated the experiments in Markham City Library, in which we included general public. Finally, we had our experiments with Maxim City summer school, which included kids and teenagers between 9 to 15 years old. In Figure~\ref{fig:vire}, a view of a sample scenario and the experiment setup with participants is shown. A video of a participant, while doing the experiment is also attached in the supplementary materials.

The process of an experiment for a participant is as follows: The participant is first asked to fill a questionnaire on his/her sociodemographic information, travel patterns and previous experiences with virtual reality (See \ref{A:var} for details of information collected). The participant is then familiarized with the VR environment for 5 minutes. After familiarization, 15 scenarios are randomly drawn from the 90 selected scenarios by the experiment design. After a scenario starts, the participant finds him or herself on the sidewalk of a marked crossroad, with vehicles approaching. The participant then have to decide to initiate his or her cross when he or she feels it is safe to do so. If a pedestrian starts crossing the street, approaching vehicles will detect them and apply braking depending on the scenario to slow down and avoid a collision. The vehicles will continued to move after the street is cleared. In this study, we solely investigate safe crosses of pedestrians. Thus, dangerous crossings with a Post Encroachment Time (PET) of less than 1.5 seconds are removed from the data and not considered in this study. For an adult participant, each scenario is conducted two times, accumulating to 30 total experiments with a total time of about 30 minutes. Our trial study showed that continuing experiments for over 30 minutes causes fatigue among participants and thus the results may be affected. For adolescent participants, each scenario is conducted once, with a total time of 15 minutes for the whole experiment. It should be noted that data from child participants are not used in this particular study as their behaviour when facing VR environment and AVs mostly involved unpredicted reactions and crossing behaviour, and we believe a different data cleaning is required on their data. A video of a young participant, while doing the experiment is provided in the supplementary materials.
\subsection{Data summary}
\begin{figure}[]
    \centering
    \includegraphics[scale=0.6]{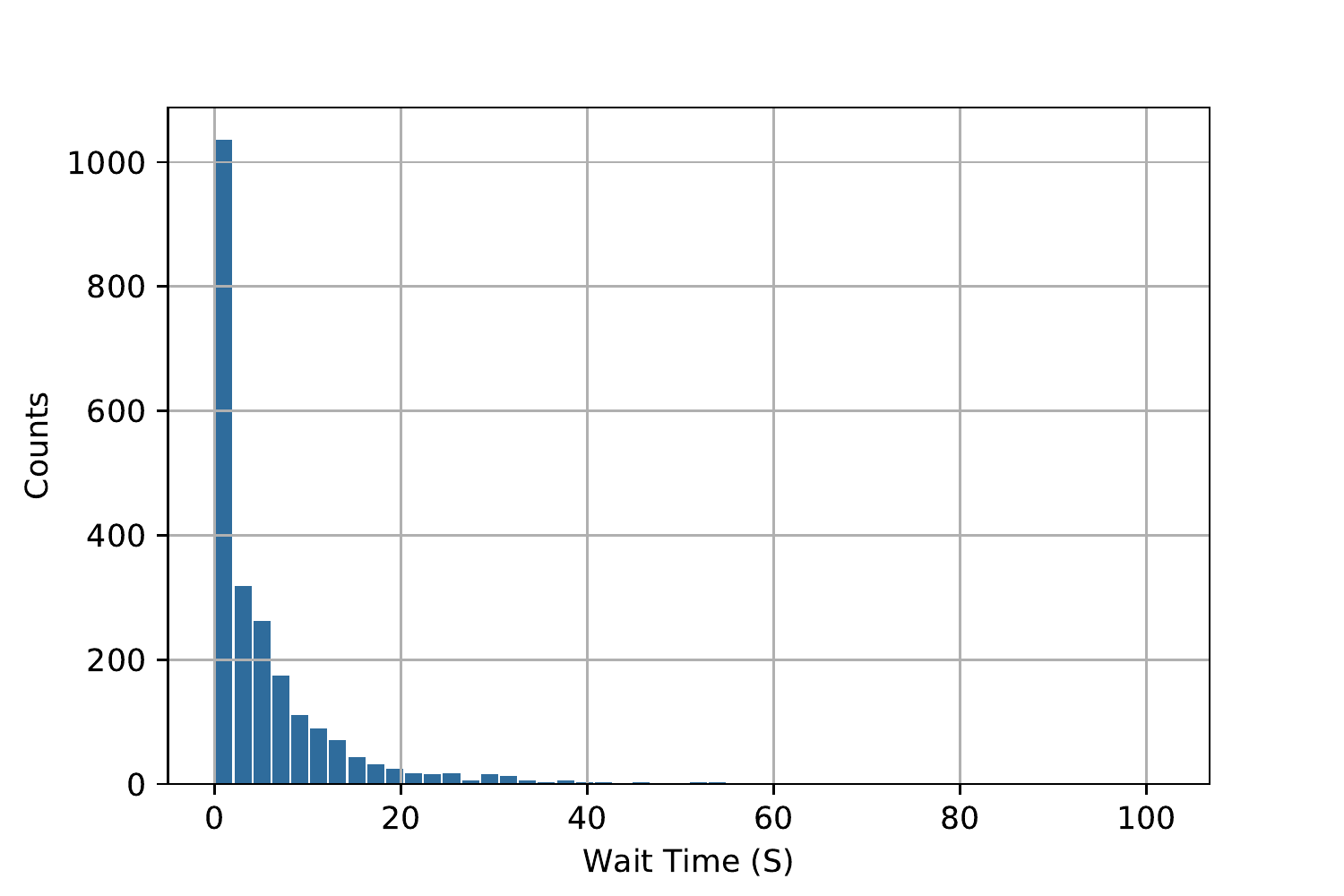}
    \caption{Wait time frequency}
    \label{fig:waitfreq}
\end{figure}

\begin{figure}[H]
    \centering
    \includegraphics
    [trim=10 10 10 10,clip,width=0.99\textwidth]{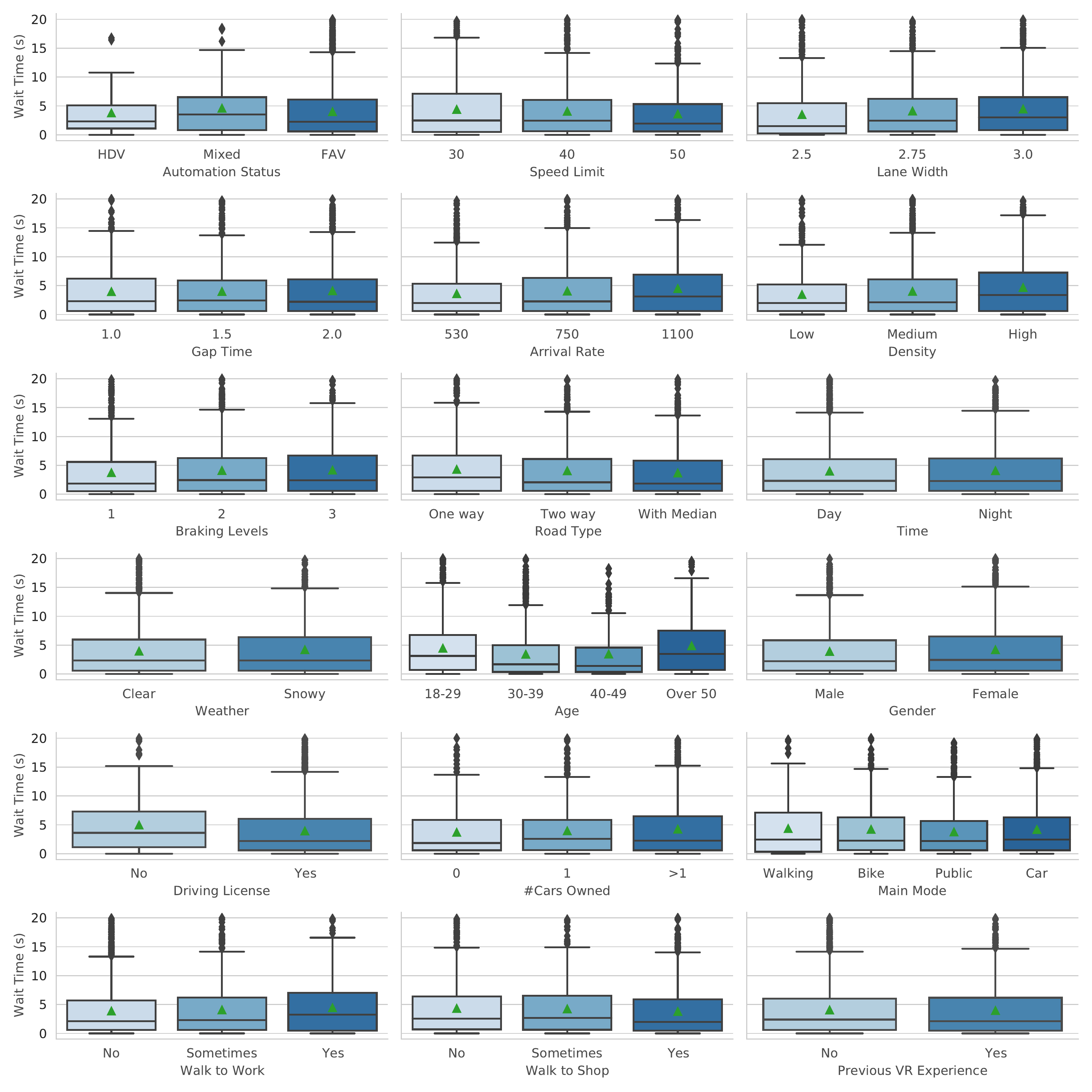}
    \caption{Comparison of wait time for different levels of covariates}
    \label{fig:comparison}
\end{figure}

Before applying feature selection algorithms, the first step in data preprocessing would be to remove highly correlated variables from the dataset. To do so, Variable Inflation Factor of covariates are calculated, which for each covariate represent the multicollinearity of that covariate. In general, a higher VIF for a covariate indicates the higher ability to predict that covariate using linear regression based on other covariates on the dataset. After removing experiments where pedestrians could not complete the crossing, or could not follow the experiments' procedures, a total of 2,291 responses were remained to be studied. Collected data are then used to estimate models for predicting pedestrian waiting time. In Figure~\ref{fig:waitfreq}, the frequencies of crosses with different wait times are provided. As it can be inferred from this figure, a majority of the crosses (54\%) occurred in the first two seconds, decaying exponentially. According to the figure, Only 6\% of the crossings took more than 20 seconds of wait time.
All quartiles of waiting time for participants, as well as the average waiting time, are depicted thorough box plots in Figure \ref{fig:comparison} for different levels of all covariates. As shown in the figure, more participants in our experiment tended to wait longer when exposed to automated vehicles, and in mixed traffic conditions participants even took longer to cross, compared to fully automated conditions. The reason might be the effect of having two different types of vehicles in these scenarios, which increases uncertainty among participants. In the second plot, waiting time over different levels of speed limit, is not as expected. Participants in our sample data waited longer in slower traffics. The reason for this behaviour can be that participants waited for a number of vehicles in scene to pass first before crossing, which takes longer for slow vehicles. Moreover, given a fixed arrival rate, higher speeds mean larger gap distances between the vehicles, which makes it easier for pedestrians to cross. Considering pedestrians tendency to decide to cross based on their distance to the vehicles, other traffic parameters like density and flow are also analyzed to complement speed limit. The next box plot reveals that in wider lane widths, participants in our experiment waited longer before crossing. No observable difference is identified in the median of minimum gap time allowed, with 1 second gap times having a slightly higher wait times compared to 2 seconds gap times. According to the arrival rate box plot, participants in more congested areas with higher traffic flows waited longer before crossing. Higher values for derived variable of vehicles' density also had a positive correlation with waiting time. Values of density are continuous in the experiment, and are categorized to three equal intervals in the figure. Three defined braking types for automated vehicles are shown next. As participants are not informed about the type of braking level before the experiments, it seems that a particular trend does not exist with them and waiting time. Two-way roads with median seemed to result in shorter waiting times compared to two-way roads with no median. Regarding environmental variables, no notable pattern based on weather conditions and time of the day is seen in the plots. 
In the age categories analyzed, two extremes of the analyzed age groups have greater waiting time, and in gender, females have slightly longer wait times.
Five next plots are relevant to travel habits of participants. Almost all of these plots show longer wait times for participants who tend to have less tendency to have active modes of transportation. Finally, participants who have prior experience in using virtual reality wait shorter on the sidewalk, as they probably feel more comfortable using the devices. Average, standard deviation and p-values for comparison on means of all the covariates depicted in box plots of Figure~\ref{fig:comparison} are presented in \ref{A:data}. It should be noted that a detailed analysis of covariates require developing survival models, which are presented in detail in section~\ref{S:results}. 

\newpage
\section{Modelling results and analysis}
\label{S:results}
Four models are developed and compared in this section: a simple binary choice model as a ground for comparison of the performance of survival models, a traditional CPH model with linear log-partial hazard function, a deep neural network-based CPH function (DCPH1), and a deep neural network-based CPH function empowered by RreliefF (DCPH2).

\subsection{Model performance}
\arash{Applying VIF before developing the models is essential to remove highly correlated variables, as models consisting of them fail to converge because of non-invertibility of singular matrices. Models are developed over a training dataset, consisting of 80\% of the data, and evaluated on a test dataset, both of which selected randomly. To measure the goodness of fit of the models, \textit{Concordance Index (C-index)} is used as the global index for comparing survival times. C-index is defined as the fraction of pairs of observations, in which the observation with higher predicted risk of starting a cross has shorter waiting time.}
\subsubsection{Binary Choice model}
As a base case for comparison of survival models' performance to other approaches, a binary choice model is developed. For every 100 milliseconds interval (the interval captured by the VR), a participant is faced with two choices: whether to initiate a cross or not. The independent variables used to develop the model are all the covariates used for developing survival models, plus a variable representing the time passed since the start of the experiment. The latter variable is needed to distinguish the instances of a single experiment from each other. Thus, a great advantage of using CPH partial hazard function, i.e. estimation of the effect of covariates regardless of the time the pedestrian has spent on the sidewalk, is missing. Applying all the data cleaning procedures and model evaluation metrics in the same way as the survival models, a C-index of 0.58 over the training set, and 0.56 over the test set is achieved.
\subsubsection{Cox Proportional Hazards model}
 After removing highly correlated variables, CPH model is developed using Python's lifeline package~\citep{cameron_davidson_pilon_2020_4002777} starting with all the remained variables and iteratively removing insignificant variables. Developing a CPH model over the train dataset using 10-fold cross validation, a C-Index of 0.57 is achieved on the test set. Details of the performance of CPH model are presented in Table~\ref{tab:compare}.
\subsubsection{Deep CPH model (DCPH)}
After applying VIF to remove highly correlated variables, variables are ranked based on their importance according to their RReliefF importance weights, before training the network. Hyperparameter $n$ is introduced as the number of covariates to be used in training the network. Other hyperparameters include: number of hidden layers, number of nodes in hidden layers, learning rate and learning rate decay for optimization, whether to use batch normalization, and dropout layer and its rate. To find the best model, random search for optimizing hyperparameters is utilized~\citep{bergstra2012random}. In Table~\ref{tab:hyper}, the optimum values for hyperparameters based on a 10-fold cross-validation over 100 epochs on training dataset are presented. 
\begin{table}[t]
\caption{Model parameters}
\centering
\begin{tabular}{|c|c|}
\hline
 \thead{\textbf{Parameter}} & \thead{\textbf{Value}} \\
\hline
     Number of Hidden Layers&3\\
     Number of Nodes in Hidden Layers&90 \\
     Batch Normalization&True\\
     Dropout Rate & 0.1\\
     Learning Rate& 0.001\\
     Learning Rate Decay& 0.001\\
     
     \hline
\end{tabular}
\label{tab:hyper}
\end{table}
Same test dataset as the one used for CPH is used again to evaluate the performance of the network over 100 epochs. Using a framework with $n=$19 top covariates as inputs, three hidden layers of 90 units each, batch normalization and a dropout rate of 0.1, a C-index of 0.64 in cross-validation and 0.62 for test set is achieved, showing an improvement of 5 percent compared to CPH.

To further assess the performance of the proposed framework, another neural-network based survival model is developed without incorporation of RreliefF and VIF (DCPH1). C-indices of all four models, along with the number of covariates incorporated in the models are presented in  Table~\ref{tab:compare}. It can be observed that in general survival models are performing better than simple binary choice models, with a slight improvement in linear CPH and greater inprovements in DCPH2. Also, DCPH2 outperforms all the other models, with 5\% and 2\% improvement over Linear CPH and DCPH1, respectively. Comparing the number of covariates used, using DCPH2, the process of removing insignificant covariates and retraining the models to find the best combination of covariates manually is no longer required. Compared to DCPH1, DCPH2 results in a better C-index with less number of covariates, meaning a less computationally expensive prediction in DCPH2.
\begin{table}[t]
\caption{Comparing the performance of the models}
\centering
\begin{tabular}{|c|c|c|c|}
\hline
 \thead{\textbf{Model}} & \thead{\textbf{Number of}\\\textbf{covariates}} & \thead{\textbf{C-index:} \\ \textbf{Validation Set}} & \thead{\textbf{C-index:} \\ \textbf{Test Set}} \\
\hline
    Binary Choice & 11(+intercept)&0.59&0.57\\
     Linear CPH&10 &0.60&0.57\\
     DCPH1&21& 0.61 & 0.60 \\
     DCPH2&19 & \textbf{0.64} &\textbf{0.62}\\ 
     \hline
\end{tabular}
\label{tab:compare}
\end{table}
\subsection{Covariates and their effects}
\subsubsection{Binary Choice model}
Table~\ref{tab:bc} shows the significant variables in the binary choice models. The large coefficient achieved for intercept, and the significance of elapsed time are the major differences in binary choice model's interpretation compared to the survival models. All the other significant variables and their effects are similar to results obtained by linear CPH (Table~\ref{tab:cox}). A positive coefficient of variable shows the effect of that variable on increasing the tendency of the participant to start crossing, whereas a negative coefficient means contribution to the decision of not to cross. Two major issues remain with the binary choice model, aside from its accuracy. First, the requirement of the variable elapsed time since the start of the experiment, makes this approach dependent on knowing the time that pedestrian has already waited on the sidewalk. CPH based models, on the other hand, rely solely on the covariates on partial hazard model, and the time parameters are left to baseline hazard function. Second issue with binary choice model is the need for discretization of wait time, and relying on decisions of cross or not, rather than the continues approach of survival models. 
\begin{table}
\caption{Binary choice model results}
\footnotesize
    \centering
    
    \begin{tabular}{|ccc|}
    \hline
         \textbf{Variable}&\textbf{Coefficient}& \textbf{p-value}  \\
    \hline
         Intercept&-3.51 & $<$0.005\\
    \hline
         Elapsed Time&-3.06&$<$0.005\\
    \hline
         Traffic Density&-0.86&$<$0.005\\
    \hline 
         Age: 30-39&0.37&$<$0.005\\
     \hline
         Lane Width&-0.34   &$<$0.005\\
    \hline
    Main Mode: Active Modes&-0.29	&$<$0.005\\
    \hline
         Road Type: Two-way with Median&  0.26&$<$0.005\\
    \hline
        Main Mode: Car&	-0.21&$<$0.005\\
    
    \hline
         Previous VR Experience& 0.21	&$<$0.005\\
    \hline
         No Cars in the Household&  0.21	&$<$0.005\\
    \hline
         Age Over 50&-0.20&0.035\\
    \hline
         Gender: Female&-0.12&0.023\\
    
    \hline
    
    \end{tabular}
    
    \label{tab:bc}
\end{table}
\subsubsection{Cox Proportional Hazards model}
Covariates used for developing CPH are listed in Table \ref{tab:cox}, along with their coefficients, Hazard Ratios and P-Values as a measure of significance. P-Value of 0.1 (90\% confidence interval) is set as the threshold of significance for selecting the covariates. The third column of the table provides \textit{Hazard Ratio}, which is defined as the ratio of the hazard rates of two conditions of a covariate. A hazard ratio of 0.84 for binary covariate \textit{Age Over 50}, for instance, means that the hazard of a pedestrian older than 50, over the hazard of other pedestrians equals 0.84, implying lower probability of crossing (longer waiting times) for pedestrians aged over 50. Hazard ratio is calculated as the exponential of coefficients in the second column. In other words, a higher coefficient, implies higher hazard ratio, and a hazard ratio greater than 1 (positive coefficient) means that the probability of a cross is higher for instances having greater values of that covariate, leading to shorter waiting times. Based on the results in Table~\ref{tab:cox}, participants who walk to do their shoppings, have previous VR experience, aged between 30 and 39, or have no cars in the household wait shorter in our sample of collected data compared to their counterparts with different values for each of the mentioned covariates. Moreover, experiments on two-way roads with median take shorter wait time compared to two-way roads with no median as the baseline variable of this covariate. On the other hand, participants who indicated their main mode of travel as cars, are aged over 50, or are females, have longer wait times. In addition, experiments in higher traffic densities and higher lane widths makes pedestrians wait longer in our sample data.

Going deeper into the effects of covariates, it can be inferred that participants who tend to have the walking habits in their life-style, i.e. walk for doing their shops and have no cars in their households, feel more comfortable crossing the streets, and have shorter waiting times. On the other hand, participants who use cars as their main mode of transportation, are expected to be less familiar with the crossing conditions and have to wait longer before initiating a cross. Considering gender and age covariates, it can be observed that females in our sample size waited longer before crossing, which is in line with some of the previous studies in the literature. The eldest age group of participants, who are over the age of 50, also tend to spend more time on the sidewalk before crossing the street. One the other hand, participants in the second youngest age group wait on the sidewalk for shorter times compared to the baseline category for age: 40 to 49. The developed model does not consider age category of between 18 and 29 as a significant variable, which can be resulted by the high heterogeneity of the participants in this age category.
Regarding variables related to geometry of the road, participants tend to wait longer before crossings roads with larger lane width. Significance of this variable can lead to the promotion of larger sizes for sidewalks and narrower lane widths. As stated earlier, out of the three road types, two-way roads with no median are settled as the baseline category of road type. Compared to the baseline category, medians help pedestrians cross with shorter waiting times.
Other significant covariate set by the scenario is traffic density on road, for which higher values lead to longer waiting times by the participants, due to the time required to find the appropriate safe gaps for crossing.
Finally, having previous VR experience helps participants of our experiment cross in shorter times. Despite having a 5-minute familiarization session with the equipment and environment before the experiments, it seems that further tutorials can be helpful to participants with no prior VR experience in future studies.

An interesting observation on the results of CPH is the insignificance of the level of traffic automation status, as well as other scenario-related variables, on participants' wait time. As different pedestrian's behaviours was expected for different scenarios, it can be inferred that linear CPH is incapable of capturing complex relationships among the covariates and wait time.   
\begin{table}
\caption{Multivariate proportional hazard model results}
\footnotesize
    \centering
    
    \begin{tabular}{|cccc|}
    \hline
         \textbf{Variable}&\textbf{Coefficient}&\textbf{Hazard Ratio}& \textbf{p-value}  \\
    \hline
         Traffic Density&-0.83& 0.43&$<$0.005\\
    \hline 
         Age: 30-39&0.32&1.38&$<$0.005\\
    \hline
         Lane Width&-0.31   &   0.73&$<$0.005\\
    \hline
         Road Type: Two-way with Median&0.22&  1.24&$<$0.005\\
    \hline
         Walk to Shopping&0.18&1.20&$<$0.005\\
    
    \hline
         Age Over 50&-0.17& 0.84&0.06\\
    
    \hline
         Previous VR Experience&0.14& 1.15&$<$0.005\\

    \hline
         No Cars in the Household&0.14&  1.15&$<$0.03\\
    \hline
         Gender: Female&-0.13& 0.87&0.01\\
    \hline
         Main Mode: Car&-0.12 & 0.88&0.03\\
    \hline

    \end{tabular}
    
    \label{tab:cox}
\end{table}

\subsubsection{Deep CPH (DCPH) model}
\begin{figure}[!h]
    \centering
    \includegraphics[scale=0.5]{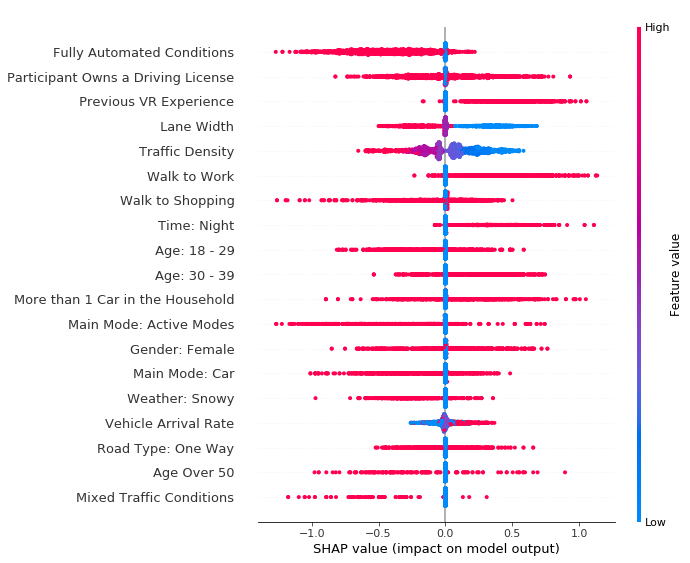}
    \caption{Plot summary of the effects of all the covariates based on the SHAP Values}
    \label{fig:shap}
\end{figure}

SHAP in its core, estimates the contribution of covariates to the waiting time for each instance separately. Each SHAP value, corresponds to the contribution of the covariate to log-partial hazard values, compared to when the covariate has a baseline value, called \textit{background} value in SHAP terminology~\citep{molnar2019interpretable}. Default baseline values for a covariate in the original paper are set to be equal to the average value of that covariate. In this study, we used the default values for continues covariates. However, the question arises for binary variables, in which having a real number value does not have a meaningful interpretation outside of the model. To address this issue, we set the baseline value for binary variables to zero, thus comparing the contribution of each binary covariate of an instance to cases where the same covariate has a value of zero. By doing so, SHAP value of a covariate in instances that the value of that covariate is zero is not calculated and contribution of a covariate is assessed based on SHAP values of the covariate in other instances. In Figure~\ref{fig:shap}, SHAP Values for the covariates for all the instances are depicted. Feature values are the values that covariates take, represented in the color. A dot in red means an instance with the higher than baseline value for that covariate, and blue means an instance with lower values in that covariate. As stated, the baseline value for binary covariates is set to zero. Thus, the instances with a value of zero in binary covariates are not taken into account for calculating the impact (blue dots on the central axis in Figure~\ref{fig:shap}).

To have a numerical understanding of SHAP values, average and standard deviation of SHAP values for all 19 covariates used in training the network, and over all instances, are presented in Table~\ref{tab:sh}, sorted by their importance according to SHAP values. As a rule of thumb, we consider the overall contribution of a covariate interpretable, if it has a relatively uniform effect on a major part of instances. In other words, if for a covariate, the absolute mean of SHAP values is greater than the standard deviation of the SHAP values, over all instances, we call the effect of that covariate uniform. It should be noted that non-uniformity of SHAP values does not mean insignificance of that covariate. Non-uniformity reveals that the interpretability method used could not capture the complex non-linear multi-level relationship among covariates. Intuitively, a covariate has non-uniform SHAP values when it has a wide range of SHAP values, extending over negative and positive values, as in these cases, the effects can vary based on the instance. 

Looking into SHAP values in Figure~\ref{fig:shap}, it can be observed that unlike CPH that was not able to capture the effects, automation level covariates have generally a negative contribution to log-partial hazard, leading to longer waiting times. Relatively high mean and low standard deviation of fully automated and mixed traffic conditions in Table~\ref{tab:sh} confirms the significant negative effect of these two covariates on log-partial hazard. As pedestrians feel more unfamiliar and concerned with cars with no drivers, the effects were expected, but could not be captured by linear CPH. By assuming that pedestrians have lower trust in AVs compared to human-driven vehicles, longer waiting time for pedestrians facing AVs in our experiments is aligned with the hypothesis made in~\citep{jayaraman2019pedestrian}, but contrasts their finding that a positive correlation exists between trust in AVs and average waiting time before cross. However, different participants' demographics and sample sizes, definition of trust, and different scenarios tested in the two studies need to be taken into account before driving to a conclusion.
\begin{table}
\caption{DCPH2 interpretability results, sorted by the absolute mean SHAP value}
\footnotesize
    \centering
    \begin{tabular}{|ccc|}
    \hline
    \textbf{Variable} &   \textbf{Mean SHAP Value} &  \textbf{std of SHAP Values} \\
    \hline
    Mixed Traffic Conditions & -0.535410 &  0.299655 \\
    \hline
    Previous VR Experience & 0.492931 &  0.193912\\
    \hline
    Fully Automated Conditions&-0.430101 &  0.272217\\
    \hline
    Walk to Work& 0.423793 &  0.260193 \\
    \hline
    Time: Night& 0.381095 &  0.202777 \\
    \hline
    Main Mode: Active Modes	&-0.372863 &  0.410266 \\
    \hline 
    Weather: Snowy&-0.232487 &  0.179340 \\
    \hline
    Age: 30 - 39	& 0.231225 &  0.263909 \\
    \hline
    Age Over 50&-0.224146 &  0.374447 \\
    \hline
    Main Mode: Car&-0.219562 &  0.288102 \\
    \hline
    Lane Width&-0.197134 &  0.174416 \\
    \hline
    Traffic Density&-0.189191 &  0.134376 \\
    \hline
    Age: 18 - 29	&-0.149895 &  0.262167 \\
    \hline
    More than 1 Car in the Household& 0.128925 &  0.308825 \\
    \hline
    Walk to Shopping&-0.072093 &  0.267495 \\
    \hline
    Gender: Female& 0.041379 &  0.296014 \\
    \hline
    Vehicle Arrival Rate& 0.041003 &  0.084269 \\
    \hline
    Road Type: One Way&-0.025349 &  0.201819 \\
    \hline
    Participant Owns a Driving License	&-0.009217 &  0.282812 \\
    \hline
    \end{tabular}
    \label{tab:sh}
\end{table}

Positive contribution captured for previous VR experience is in line with the results obtained in CPH, which confirms the necessity of stronger VR tutorials before experiments. Similar to CPH, wider lane width and higher traffic density both have negative effect on log-partial hazard, implying longer waiting times. Their mean and standard deviation confirm the uniformity of SHAP values for these two covariates. The effect of lane width is in line with~\citep{rasouli2017agreeing}, where authors, using real video data from urban and suburban roads, report pedestrians paying more attention before crossing wider streets. The reported effects of traffic parameters on pedestrian behaviour in previous studies also confirm the negative effect of traffic density on waiting time~\citep{schmidt2009pedestrians,ishaque2008behavioural}. \cite{schmidt2009pedestrians}, for instance, suggest speed and density as important parameters for pedestrians crossing decision, with pedestrians using distance of the vehicles for making crossing decisions.
Six of the used covariates are relevant to walking habits: Walk to work and shopping, main transportation mode of cars and active modes, having driving licence and having more than 1 car in the house. Among them, only walking to work appears to be interpretable based on mean and standard deviation of SHAP values, which shows shorter waiting time for participants who indicated that they walk to commute to their works. Similar results are obtained in \citep{hamed2001analysis} with analysing wait time of pedestrians in divided and undivided streets. Based on their study, pedestrians owning cars have longer wait times due to their greater perception of risk. 
Environmental covariates, on the other hand, are added to the model and appear to have uniform effect. Results showed that participants crossing in simulated snowy scenarios, have longer wait times, which may be traced to poor sight distance. Previous studies confirm this observation, where bad weather conditions are associated with a negative effect on pedestrians' perception of speed, and being more conservative~\citep{rasouli2019autonomous,sun2015estimation,harrell1991factors}. On the other hand, in the night scenarios, participants tended to have shorter wait times, which is not consistent with the supposed poor sight distance in the simulation. Poor sight distance at nights has also been associated with riskier decisions ~\citep{rasouli2019autonomous}, which might justify the shorter wait times of pedestrians. However, in our study we solely focus on safe crossings and it is not clear whether we can assume crosses with shorter wait times more \textit{risky}. The contribution of night time to shorter wait time in our study may be due to the fact that in night scenarios simulated, unlike snowy simulated scenarios, the sight observed by participants was not limited and changes were only made to the color of the sky to project the mental effect of night. This can be addressed in future data collection campaigns. 
The effect of road type covariate did not appear to be interpretable over all instances, and vehicle arrival rate, although not one of CPH covariates, appeared to be one of the selected covariates by DCPH2, without captured uniform contribution for all instances. 

\arash{Interactions with other covariates may be investigated to further analyze the effects of covariates with non-uniform contributions. Numerous possible combinations can be defined and investigated. In this study, however, we limit our analysis to two-level combinations, and more complex combinations are left as a future direction. Table~\ref{tab:int} shows the two-level interactive effect of the investigated covariates. All the provided combinations have appeared to be significant based on the values of the mean and standard deviation of their SHAP values. The first column represents the variable upon which the combination is conditioned, and the second column of the table shows the variable that its effect is investigated under the condition variable. To better understand the effect of age and gender, combined effects of some covariates conditioned upon three relevant covariates to age and gender are investigated. Gender is arguably one of the determining factors in explaining pedestrians' behaviour~\cite{rasouli2019autonomous,hamed2001analysis}. The differences in crossing behaviour of males and females are traced back to differences in motives~\citep{yagil2000beliefs}, attention to environment~\citep{tom2011gender}, and females being more cautious~\citep{holland2007effect}. However, this trend in general is not observed in our modeling results. Part of this inconsistency might be explained by the use of VR in our experiment, where details are not as elaborate as real life, risk involved is less and laws are not as strict as real streets. Interaction of gender with some other variables, however, suggest gender contribution to wait time in some scenarios: In mixed traffic conditions, females wait longer. The pattern exists in fully automated conditions, but the contribution is not uniform and thus we have not included it in the table. Among those participants whose main mode of transportation is their car, females appear to be more conservative. In one-way roads, females also tend to wait longer compared to men. Gender appeared to be insignificant in different weather conditions and times of the day. Results also show that having strong walking habits helps both younger and older age groups cross in shorter times. On the other hand, although participants of age over 50 wait longer in the presence of automated vehicles in the experiment, younger participants do not seem to be affected significantly by this factor. Similarly, among participants aged over 50, waiting in snowy weather takes longer. Similar trends occur for female participants, as a lack of walking habits, presence of AVs and snowy weather conditions make the wait times before cross take longer. Also among females, having an age of over 50 contributes to longer wait times. Generally, studies on the effect of age suggest elderly population are more cautious~\citep{sun2003modeling,harrell1991precautionary,hamed2001analysis}. Although age does not have a uniform contribution in our case, having an age of over 50 in poor sight distances, and for females, is shown to be a contributing factor.  Effects of walking habits, automation condition of the vehicles on the road, age and snowy weathers during night scenarios follow an expected pattern. When night scenario are accompanied by snowy weathers and thus poor sight distances, the waiting times before crosses are increased. Also, elderly people are behaving more conservatively in night scenarios as opposed to other age brackets in our experiment. Higher traffic densities and wider lane widths were both associated with longer wait times based on the single-level analysis in Table~\ref{tab:sh}. However, having walking habits appear to be effective in decreasing the wait time among high density scenarios and wide lane width scenarios. Fully automated or mixed traffic conditions also contribute to longer waiting times, even among high traffic densities and wide lane width scenarios.}
\begin{table}
\caption{\arash{DCPH2 interaction variables results}}
\footnotesize
    \centering
    \begin{tabular}{|ccc|}
    \hline
    \textbf{Condition Variable} &   \textbf{Variable} &  \textbf{Mean SHAP Value} \\
    \hline
    Mixed Traffic Condition & \multirow{3}{*}{Gender: Female} &-0.100629 \\ Main Mode: Car& &-0.263754\\Road Type: One Way && -0.229365 \\
    \hline
    \hline
    \multirow{2}{*}{Age 18 - 29} & Walk to Work &  0.437002  \\
    &Main Mode: Car &  -0.377411\\
    \hline
    \hline
    \multirow{4}{*}{Age Over 50} & Walk to Work &  0.668556 \\
    
     & Weather: Snowy&  -0.377821 \\
    
    &Fully Automated Conditions &  -0.614000\\
    &Mixed Traffic Conditions &  -0.630974\\
    \hline
    \hline
    \multirow{5}{*}{Gender: Female} & Walk to Work &  0.434888 \\
        &Main Mode: Car  &  -0.497206\\
     & Weather: Snowy&  -0.197193 \\
    
    &Fully Automated Conditions &  -0.331376\\
    &Mixed Traffic Conditions &  -0.467947\\
    &Age Over 50  & -0.385490\\
    \hline
    \hline
     \multirow{5}{*}{Time: Night} & Walk to Work &  0.315519 \\
    &Fully Automated Conditions &  -0.571182\\
    &Mixed Traffic Conditions&  -0.501431\\
    &Weather: Snowy  & -0.234872\\
    &Age Over 50 & -0.200602\\

    \hline
    \hline
     \multirow{5}{*}{High Traffic Density} & Walk to Work &  0.371923 \\
    
    &Fully Automated Conditions &  -0.463269\\
    &Mixed Traffic Conditions&  -0.752063\\
    &Weather: Snowy  & -0.275998\\

    \hline
    \hline
    \multirow{3}{*}{Wide Lane Width} & Walk to Work &  0.449537 \\
    &Fully Automated Conditions &  -0.344592\\
    &Mixed Traffic Conditions&  -0.535366\\
    \hline
    \hline
    \end{tabular}
    \label{tab:int}
\end{table}

\newpage
\section{Discussion and policy implications}
\label{S:discuss}
There are several key practical implications and useful policy recommendations that can be derived from the virtual reality experiments, models, and results of this study, which can be of use to urban planners, policy makers, manufacturers of automated vehicles, and researchers seeking to implement virtual reality and data-driven modelling in their work.

During the VR data collection campaign, out of 180 participants, a total of 113 adults participated who were able to complete at least one of their 30 assigned scenarios. However, after data cleaning, 2,291 responses remained which account for 67.6\% of our expectation on the number of total responses (113 $\times$ 30). Aside from dangerous crossings with PETs less than 1.5 seconds that we removed from the data for the purpose of the study, two major reasons contribute to this inefficiency in data collection were:
\begin{itemize}
    \item Participants feeling motion sickness after some attempts: One of the major reasons of failing to use a participants' task data was motion sickness. Despite priorly asking of participants' background health issues, not all the participants were aware of their reaction in VR. Widespread development of VR can help ease this issue in future studies with participants better knowing their reactions and technology building smoother environments.
    \item Participants not following guidelines: Despite the training session dedicated to teach participants on how to do the tasks, some experiments were removed due to their failure to do so. Easier to follow guidelines and continues reminders during the experiment can help to prevent this issue in the future. 
\end{itemize}

Regarding participants' experience in VR, extra attention should be paid to children and senior participants in terms of training and educational programs, and their comfort during the experiments. Moreover, providing facilities for people with special needs should be taken into account in future research to make experiments more inclusive. Out of the 180 participants recruited for the experiments, 20 who were children and senior participants, failed to complete any of their tasks. Although we had one person using wheelchair as a participant, more accessibility services should be provided in future to increase the participation of people with different disabilities. As per our observations, some children tended to see the virtual reality tools as gaming platforms. Irrational crossing behaviours to explore the 3D environment, followed by their loss of interest in continuing the experiment after observation of different elements of the environment, led to their failure in completing the experiment tasks. The virtual vehicles were designed such that they could foresee if the pedestrian is in danger and would apply braking when needed. Some teenage participants realized that and started to play with the virtual vehicles. They would come in front of an vehicles so it would stop. Participant would then move back, and virtual vehicle would sense this action and start moving, at which point, the participant would move in front of the vehicle again causing it to stop, repeating such actions several times. This behaviour is expected to happen when the fully automated vehicles are operating on urban roads, especially in lower speeds, resulting in disruption of the upstream traffic. To avoid such behaviour, VR based education and training is needed. Municipalities can also look into proactive regulations to minimize such disruptions. As for senior participants, their inconvenience while using VR headsets, inability to ambulate independently without mobility devices such as wheelchair, and feeling nauseated and fatigued while interacting with 3D immersive environment were identified in our observations as the main cause of them not completing the tasks.

Interpretation of our model results can be useful for practical implications for urban decision makers and car manufacturers. Our results show that participants tend to be more conservative and cautious in the presence of automated vehicles. Either in fully automated or mixed automated and human-driven conditions of traffic, participants waited longer compared to solely human-driven conditions. \arash{Longer wait time in the presence of AVs is in particular bolder for participants aged over 50, and in congested areas with higher traffic densities. Longer wait before crossing an environment with automated vehicles in it is intuitive, and was expected considering the unfamiliarity of pedestrian to vehicles with no drivers. This is of importance considering that shorter waiting times in safe crossings of pedestrians imply more trust and confidence confronting the vehicles. Thus, predictive models for pedestrian wait time should take into account the differences of pedestrian behaviour facing vehicles with no drivers. Also, considering the unfamiliarity of pedestrians with automated vehicles in the first years of introduction of AVs, this observation requires attention of city planners and decision makers to make modifications needed to enhance pedestrians' crossing experience. Demography of the area, traffic parameters, and road geometry should all be considered in the modifications to be made.} Nationwide educational training programs should be practiced before the transition to automated environments, to familiarize pedestrians with new dynamics of the city. Immersive and dynamic virtual reality technology can play an important role in such programs, to ensure providing safe and naturalistic experience to users. Manufacturers should consider alternative ways to improve the communications and quality of interaction between the driver and pedestrians in automated environments. Some manufacturers, like drive.ai, have introduced screens on their vehicles that can give visual cues to pedestrians. Similar to \cite{fridman2017walk}, \cite{chang2017eyes}, \cite{de2019external}, etc., such new treatments can be systematically optimized and their effectiveness can be studied in VIRE environment.
Based on results of our study, narrower lane widths, lower traffic densities, and better sight distances are also revealed to be affecting parameters in pedestrians' crossing behaviors, leading to shorter wait times. To the best of our knowledge, the effect of road design parameters in the context of automated environments, i.e. lane width and existence of median, are not investigated in the literature despite their importance in crossing behaviour of pedestrians in traditional studies. Several studies both in the automated environments~\citep{rasouli2019autonomous} and human-driven environments~\citep{sun2015estimation} have investigated the effect of environmental variable, with which our results are in line. The effect of traffic parameters, in various forms, has also been explored in traditional studies of pedestrians~\citep{schmidt2009pedestrians,ishaque2008behavioural}, and we show the same pattern exists in the new context. Wider and more comfortable sidewalks, narrower lane widths, enhanced lighting equipment, and incorporation of pedestrian-to-vehicle communication technologies are some of the solutions that can be implemented before diving into future automated urban areas. Our study also reveals that having frequent walking habits positively affects the crossing experience. In the context of automated vehicles, several studies had addressed cultural differences as a contributing factor to pedestrians crossing behaviour. While this observation might hold true, using more individualized parameters like walking habits might help better address the cultural differences, particularly in cities with diverse population. Promoting active modes and developing more pedestrian-friendly infrastructures could lead to better crossing experience on the streets of the future. \arash{Multi-level analysis in our study also showed that having walking habits leads to shorter waiting times even among groups that are conventionally known to be more conservative, i.e. females and elderly people. Gender alone did not appear to have a significant contribution in our study, except for in automated and mixed traffic conditions, one-way roads, and among participants who use car as their main mode of transportation. Gender has traditionally been a significant variable in pedestrians' crossing behaviour. The limited contribution of gender in our study might be due to the virtual nature of our experiments, in which perception of risk is inevitably lower that real roads. Our results also showed that in poor sight distances, as well as among female participants, participants aged over 50 tend to have longer waiting times, which reveals the necessity of a special attention to this group.}

\section{Conclusion and future direction}
\label{S:Con}
In this paper, we investigated factors affecting pedestrians' wait time before crossing unsignalized crosswalks. With the upcoming revolutionary technologies on the roads, it is of vital importance to re-think and re-asses pedestrian behaviour in the presence of these unprecedented technologies. We tried to fill the existing research gap of investigating factors that appear to be significant contributions to pedestrians' crossing behaviour. Studying new technologies require new tools, thus, a virtual reality-based immersive and dynamic experiment was introduced in this study to obtain high-dimensional data from customizable scenarios in a safe way. Over a period of five months, a total of 180 people, selected in a heterogeneous and inclusive way, participated in our experiments. Participants were asked to cross an unsignalized crosswalk in various scenarios, incorporating speed limit, lane width, vehicles' arrival rate, road type, level of automation, braking system of cars, minimum gap between vehicles, weather conditions, and time of the day. Moreover, demographic information and travel habits of participants were collected using questionnaires before each experiment.

To analyze the effects of different covariates on pedestrians' wait time, survival analysis models were developed in this study. A traditional Cox Proportional Hazards (CPH) model as the baseline model, and a neural network based CPH model empowered by feature selection and interpretability methods were developed and trained in this study. Using an interpretable framework helps us capture nonlinearities among high-dimensional data, resulting in better goodness of fit achieved compared to the baseline model. Moreover, our framework outperforms simple neural-network based CPH models by following an embedded algorithm for systematic feature selection. Using a game theoretic-based interpretability method, we seek to replace traditional CPH methods by fulfilling their power in capturing the effects of covariates on baseline hazards function.

Based on the interpretability results of our study, we suggested some key practical implications and policy recommendations. Our results showed that pedestrians wait longer before crossing in the presence of automated vehicles. Wider lane widths, higher traffic densities, poor sight distances, and lack of walking habits in pedestrians found to be the other contributing factors to pedestrian wait time. Widespread educational campaigns before introduction of AVs on streets, enhanced safety measures on AVs, promoting active transportation modes, incorporating pedestrian-friendly infrastructures on streets and using pedestrian-to-vehicle communications are some of the possible solutions for future urban areas.
In conclusion, this study tries to contribute to the current literature in three general aspects:
\begin{itemize}
    \item Utilizing immersive virtual reality tools for relatively large-scale data collection, backed by a systematic design of experiment to optimize the information that can be inferred from the data. To the best of our knowledge, our VR data collection campaign is one of the biggest such campaigns for pedestrians studies. 
    \item Developing a neural-network-based survival model, to analyze the effects of different parameters on pedestrians' wait time. Incorporating neural network within a CPH model, we achieved an improvement in accuracy by 5\% compared to linear CPH models. 
    \item Using SHAP, as a modern game theoretic-based approach for neural network interpretation. Interpreting neural networks is gaining more popularity in recent years, in attempts to transfer black-box models to explainable models that can be used for policy and decision making. Limited number of studies have touched model interpretability in transportation.  
\end{itemize}

Our study was not without limitations. In terms of the data collection, as stated in the paper, participants training procedure can be enhanced to decrease the effect of previous VR experience on the performance. Providing accessibility services for people with disabilities to have more inclusive data collection is another important direction that needs to be followed. Enhancing quality of scenarios, including a broader investigation of lighting/weather condition, and different types of crossing, e.g. signalized and unmarked crossings, can be important directions to follow. Moreover, the effect of AV can be better validated by having an actual human driver for HDV scenarios, and conducting post-experiment surveys from participants to validate if they could distinguish between the two types of vehicles. Other methods of design of experiment may be also tried to compare the performance of the proposed D-Optimal design introduced in this study. Regarding the model, other deep networks can be utilized within or independent of the CPH model. Predicting pedestrian wait time using developed hazard functions can be addressed in future studies, which requires developing a time-dependant baseline hazard function. Current findings can be useful in the development of new audio, visual or direct vehicle-to-pedestrian communication methods that can make the pedestrian crossing more convenient. The effectiveness of such methods can also be tested in the virtual reality environment. Our paper is a part of an on-going research on interactions of humans and automated vehicle, which aims to develop prediction and training tools to develop socially-aware automated vehicles. Other directions of research, such as investigating crash and incidents, safety measures, etc. can be studied in future research using more comprehensive data available on the topics, or developing crash related scenarios in virtual reality.
\bibliographystyle{elsarticle-harv}

\begin{thebibliography}{73}
\expandafter\ifx\csname natexlab\endcsname\relax\def\natexlab#1{#1}\fi
\providecommand{\url}[1]{\texttt{#1}}
\providecommand{\href}[2]{#2}
\providecommand{\path}[1]{#1}
\providecommand{\DOIprefix}{doi:}
\providecommand{\ArXivprefix}{arXiv:}
\providecommand{\URLprefix}{URL: }
\providecommand{\Pubmedprefix}{pmid:}
\providecommand{\doi}[1]{\href{http://dx.doi.org/#1}{\path{#1}}}
\providecommand{\Pubmed}[1]{\href{pmid:#1}{\path{#1}}}
\providecommand{\bibinfo}[2]{#2}
\ifx\xfnm\relax \def\xfnm[#1]{\unskip,\space#1}\fi
\bibitem[{Alelyani et~al.(2018)Alelyani, Tang and Liu}]{alelyani2018feature}
\bibinfo{author}{Alelyani, S.}, \bibinfo{author}{Tang, J.},
  \bibinfo{author}{Liu, H.}, \bibinfo{year}{2018}.
\newblock \bibinfo{title}{Feature selection for clustering: A review}, in:
  \bibinfo{booktitle}{Data Clustering}. \bibinfo{publisher}{Chapman and
  Hall/CRC}, pp. \bibinfo{pages}{29--60}.
\bibitem[{Animesh et~al.(2011)Animesh, Pinsonneault, Yang and
  Oh}]{animesh2011odyssey}
\bibinfo{author}{Animesh, A.}, \bibinfo{author}{Pinsonneault, A.},
  \bibinfo{author}{Yang, S.B.}, \bibinfo{author}{Oh, W.}, \bibinfo{year}{2011}.
\newblock \bibinfo{title}{An odyssey into virtual worlds: exploring the impacts
  of technological and spatial environments on intention to purchase virtual
  products}.
\newblock \bibinfo{journal}{Mis Quarterly} , \bibinfo{pages}{789--810}.
\bibitem[{Atkinson et~al.(2007)Atkinson, Donev and
  Tobias}]{atkinson2007optimum}
\bibinfo{author}{Atkinson, A.}, \bibinfo{author}{Donev, A.},
  \bibinfo{author}{Tobias, R.}, \bibinfo{year}{2007}.
\newblock \bibinfo{title}{Optimum experimental designs, with SAS}.
  volume~\bibinfo{volume}{34}.
\newblock \bibinfo{publisher}{Oxford University Press}.
\bibitem[{Bergstra and Bengio(2012)}]{bergstra2012random}
\bibinfo{author}{Bergstra, J.}, \bibinfo{author}{Bengio, Y.},
  \bibinfo{year}{2012}.
\newblock \bibinfo{title}{Random search for hyper-parameter optimization}.
\newblock \bibinfo{journal}{Journal of Machine Learning Research}
  \bibinfo{volume}{13}, \bibinfo{pages}{281--305}.
\bibitem[{Bhagavathula et~al.(2018)Bhagavathula, Williams, Owens and
  Gibbons}]{bhagavathula2018reality}
\bibinfo{author}{Bhagavathula, R.}, \bibinfo{author}{Williams, B.},
  \bibinfo{author}{Owens, J.}, \bibinfo{author}{Gibbons, R.},
  \bibinfo{year}{2018}.
\newblock \bibinfo{title}{The reality of virtual reality: A comparison of
  pedestrian behavior in real and virtual environments}, in:
  \bibinfo{booktitle}{Proceedings of the Human Factors and Ergonomics Society
  Annual Meeting}, \bibinfo{organization}{SAGE Publications Sage CA: Los
  Angeles, CA}. pp. \bibinfo{pages}{2056--2060}.
\bibitem[{Bloomberg and Aspen(2017)}]{kinaSUR}
\bibinfo{author}{Bloomberg, P.}, \bibinfo{author}{Aspen, I.},
  \bibinfo{year}{2017}.
\newblock \bibinfo{title}{{Barriers to city AV reports}}.
\newblock
  \bibinfo{howpublished}{\url{http://avsincities.bloomberg.org/global-atlas/about}}.
\newblock \bibinfo{note}{Online; accessed 01 October 2017}.
\bibitem[{Cavalcante and Roorda(2011)}]{cavalcante2011bayesian}
\bibinfo{author}{Cavalcante, R.}, \bibinfo{author}{Roorda, M.},
  \bibinfo{year}{2011}.
\newblock \bibinfo{title}{Bayesian approach for identifying efficient
  stated-choice survey designs with reduced prior information}.
\newblock \bibinfo{journal}{Transportation Research Record: Journal of the
  Transportation Research Board} , \bibinfo{pages}{38--46}.
\bibitem[{Chang et~al.(2017)Chang, Toda, Sakamoto and Igarashi}]{chang2017eyes}
\bibinfo{author}{Chang, C.M.}, \bibinfo{author}{Toda, K.},
  \bibinfo{author}{Sakamoto, D.}, \bibinfo{author}{Igarashi, T.},
  \bibinfo{year}{2017}.
\newblock \bibinfo{title}{Eyes on a car: an interface design for communication
  between an autonomous car and a pedestrian}, in:
  \bibinfo{booktitle}{Proceedings of the 9th International Conference on
  Automotive User Interfaces and Interactive Vehicular Applications}, pp.
  \bibinfo{pages}{65--73}.
\bibitem[{Clamann et~al.(2017)Clamann, Aubert and
  Cummings}]{clamann2017evaluation}
\bibinfo{author}{Clamann, M.}, \bibinfo{author}{Aubert, M.},
  \bibinfo{author}{Cummings, M.L.}, \bibinfo{year}{2017}.
\newblock \bibinfo{title}{Evaluation of vehicle-to-pedestrian communication
  displays for autonomous vehicles}.
\newblock \bibinfo{type}{Technical Report}.
\bibitem[{Cox(1972)}]{cox1972regression}
\bibinfo{author}{Cox, D.R.}, \bibinfo{year}{1972}.
\newblock \bibinfo{title}{Regression models and life-tables}.
\newblock \bibinfo{journal}{Journal of the Royal Statistical Society: Series B
  (Methodological)} \bibinfo{volume}{34}, \bibinfo{pages}{187--202}.
\bibitem[{Das et~al.(2005)Das, Manski and Manuszak}]{das2005walk}
\bibinfo{author}{Das, S.}, \bibinfo{author}{Manski, C.F.},
  \bibinfo{author}{Manuszak, M.D.}, \bibinfo{year}{2005}.
\newblock \bibinfo{title}{Walk or wait? an empirical analysis of street
  crossing decisions}.
\newblock \bibinfo{journal}{Journal of Applied Econometrics}
  \bibinfo{volume}{20}, \bibinfo{pages}{529--548}.
\bibitem[{Das and Tsapakis(2019)}]{das2019interpretable}
\bibinfo{author}{Das, S.}, \bibinfo{author}{Tsapakis, I.},
  \bibinfo{year}{2019}.
\newblock \bibinfo{title}{Interpretable machine learning approach in estimating
  traffic volume on low-volume roadways}.
\newblock \bibinfo{journal}{International Journal of Transportation Science and
  Technology} .
\bibitem[{Davidson-Pilon et~al.(2020)Davidson-Pilon, Kalderstam, Jacobson, sean
  reed, Kuhn, Zivich, Williamson, AbdealiJK, Datta, Fiore-Gartland, Parij,
  WIlson, Gabriel, Moneda, Moncada-Torres, Stark, Gadgil, Jona, Singaravelan,
  Besson, Peña, Anton, Klintberg, GrowthJeff, Noorbakhsh, Begun, Kumar,
  Hussey, Golland and jlim13}]{cameron_davidson_pilon_2020_4002777}
\bibinfo{author}{Davidson-Pilon, C.}, \bibinfo{author}{Kalderstam, J.},
  \bibinfo{author}{Jacobson, N.}, \bibinfo{author}{sean reed},
  \bibinfo{author}{Kuhn, B.}, \bibinfo{author}{Zivich, P.},
  \bibinfo{author}{Williamson, M.}, \bibinfo{author}{AbdealiJK},
  \bibinfo{author}{Datta, D.}, \bibinfo{author}{Fiore-Gartland, A.},
  \bibinfo{author}{Parij, A.}, \bibinfo{author}{WIlson, D.},
  \bibinfo{author}{Gabriel}, \bibinfo{author}{Moneda, L.},
  \bibinfo{author}{Moncada-Torres, A.}, \bibinfo{author}{Stark, K.},
  \bibinfo{author}{Gadgil, H.}, \bibinfo{author}{Jona},
  \bibinfo{author}{Singaravelan, K.}, \bibinfo{author}{Besson, L.},
  \bibinfo{author}{Peña, M.S.}, \bibinfo{author}{Anton, S.},
  \bibinfo{author}{Klintberg, A.}, \bibinfo{author}{GrowthJeff},
  \bibinfo{author}{Noorbakhsh, J.}, \bibinfo{author}{Begun, M.},
  \bibinfo{author}{Kumar, R.}, \bibinfo{author}{Hussey, S.},
  \bibinfo{author}{Golland, D.}, \bibinfo{author}{jlim13},
  \bibinfo{year}{2020}.
\newblock \bibinfo{title}{Camdavidsonpilon/lifelines: v0.25.4}.
\newblock \URLprefix \url{https://doi.org/10.5281/zenodo.4002777},
  \DOIprefix\doi{10.5281/zenodo.4002777}.
\bibitem[{De~Clercq et~al.(2019)De~Clercq, Dietrich, N{\'u}{\~n}ez~Velasco,
  De~Winter and Happee}]{de2019external}
\bibinfo{author}{De~Clercq, K.}, \bibinfo{author}{Dietrich, A.},
  \bibinfo{author}{N{\'u}{\~n}ez~Velasco, J.P.}, \bibinfo{author}{De~Winter,
  J.}, \bibinfo{author}{Happee, R.}, \bibinfo{year}{2019}.
\newblock \bibinfo{title}{External human-machine interfaces on automated
  vehicles: effects on pedestrian crossing decisions}.
\newblock \bibinfo{journal}{Human factors} \bibinfo{volume}{61},
  \bibinfo{pages}{1353--1370}.
\bibitem[{Deb(2017)}]{deb2017pedestrians}
\bibinfo{author}{Deb, S.}, \bibinfo{year}{2017}.
\newblock \bibinfo{title}{Pedestrians' Receptivity Toward Fully Autonomous
  Vehicles}.
\newblock \bibinfo{publisher}{Mississippi State University}.
\bibitem[{Deb et~al.(2017)Deb, Carruth, Sween, Strawderman and
  Garrison}]{deb2017efficacy}
\bibinfo{author}{Deb, S.}, \bibinfo{author}{Carruth, D.W.},
  \bibinfo{author}{Sween, R.}, \bibinfo{author}{Strawderman, L.},
  \bibinfo{author}{Garrison, T.M.}, \bibinfo{year}{2017}.
\newblock \bibinfo{title}{Efficacy of virtual reality in pedestrian safety
  research}.
\newblock \bibinfo{journal}{Applied ergonomics} \bibinfo{volume}{65},
  \bibinfo{pages}{449--460}.
\bibitem[{Faiola et~al.(2013)Faiola, Newlon, Pfaff and
  Smyslova}]{faiola2013correlating}
\bibinfo{author}{Faiola, A.}, \bibinfo{author}{Newlon, C.},
  \bibinfo{author}{Pfaff, M.}, \bibinfo{author}{Smyslova, O.},
  \bibinfo{year}{2013}.
\newblock \bibinfo{title}{Correlating the effects of flow and telepresence in
  virtual worlds: Enhancing our understanding of user behavior in game-based
  learning}.
\newblock \bibinfo{journal}{Computers in Human Behavior} \bibinfo{volume}{29},
  \bibinfo{pages}{1113--1121}.
\bibitem[{Faraggi and Simon(1995)}]{faraggi1995neural}
\bibinfo{author}{Faraggi, D.}, \bibinfo{author}{Simon, R.},
  \bibinfo{year}{1995}.
\newblock \bibinfo{title}{A neural network model for survival data}.
\newblock \bibinfo{journal}{Statistics in medicine} \bibinfo{volume}{14},
  \bibinfo{pages}{73--82}.
\bibitem[{Farooq et~al.(2018)Farooq, Cherchi and Sobhani}]{farooq2018virtual}
\bibinfo{author}{Farooq, B.}, \bibinfo{author}{Cherchi, E.},
  \bibinfo{author}{Sobhani, A.}, \bibinfo{year}{2018}.
\newblock \bibinfo{title}{Virtual immersive reality for stated preference
  travel behavior experiments: A case study of autonomous vehicles on urban
  roads}.
\newblock \bibinfo{journal}{Transportation Research Record} ,
  \bibinfo{pages}{0361198118776810}.
\bibitem[{Fisher et~al.(2018)Fisher, Rudin and Dominici}]{fisher2018all}
\bibinfo{author}{Fisher, A.}, \bibinfo{author}{Rudin, C.},
  \bibinfo{author}{Dominici, F.}, \bibinfo{year}{2018}.
\newblock \bibinfo{title}{All models are wrong but many are useful: Variable
  importance for black-box, proprietary, or misspecified prediction models,
  using model class reliance}.
\newblock \bibinfo{journal}{arXiv preprint arXiv:1801.01489} .
\bibitem[{Fridman et~al.(2017)Fridman, Mehler, Xia, Yang, Facusse and
  Reimer}]{fridman2017walk}
\bibinfo{author}{Fridman, L.}, \bibinfo{author}{Mehler, B.},
  \bibinfo{author}{Xia, L.}, \bibinfo{author}{Yang, Y.},
  \bibinfo{author}{Facusse, L.Y.}, \bibinfo{author}{Reimer, B.},
  \bibinfo{year}{2017}.
\newblock \bibinfo{title}{To walk or not to walk: Crowdsourced assessment of
  external vehicle-to-pedestrian displays}.
\newblock \bibinfo{journal}{arXiv preprint arXiv:1707.02698} .
\bibitem[{Friedman(2001)}]{friedman2001greedy}
\bibinfo{author}{Friedman, J.H.}, \bibinfo{year}{2001}.
\newblock \bibinfo{title}{Greedy function approximation: a gradient boosting
  machine}.
\newblock \bibinfo{journal}{Annals of statistics} ,
  \bibinfo{pages}{1189--1232}.
\bibitem[{Goldstein et~al.(2015)Goldstein, Kapelner, Bleich and
  Pitkin}]{goldstein2015peeking}
\bibinfo{author}{Goldstein, A.}, \bibinfo{author}{Kapelner, A.},
  \bibinfo{author}{Bleich, J.}, \bibinfo{author}{Pitkin, E.},
  \bibinfo{year}{2015}.
\newblock \bibinfo{title}{Peeking inside the black box: Visualizing statistical
  learning with plots of individual conditional expectation}.
\newblock \bibinfo{journal}{Journal of Computational and Graphical Statistics}
  \bibinfo{volume}{24}, \bibinfo{pages}{44--65}.
\bibitem[{Hagenauer and Helbich(2017)}]{hagenauer2017comparative}
\bibinfo{author}{Hagenauer, J.}, \bibinfo{author}{Helbich, M.},
  \bibinfo{year}{2017}.
\newblock \bibinfo{title}{A comparative study of machine learning classifiers
  for modeling travel mode choice}.
\newblock \bibinfo{journal}{Expert Systems with Applications}
  \bibinfo{volume}{78}, \bibinfo{pages}{273--282}.
\bibitem[{Hamed(2001)}]{hamed2001analysis}
\bibinfo{author}{Hamed, M.M.}, \bibinfo{year}{2001}.
\newblock \bibinfo{title}{Analysis of pedestrians’ behavior at pedestrian
  crossings}.
\newblock \bibinfo{journal}{Safety science} \bibinfo{volume}{38},
  \bibinfo{pages}{63--82}.
\bibitem[{Harrell(1991a)}]{harrell1991factors}
\bibinfo{author}{Harrell, W.A.}, \bibinfo{year}{1991}a.
\newblock \bibinfo{title}{Factors influencing pedestrian cautiousness in
  crossing streets}.
\newblock \bibinfo{journal}{The Journal of Social Psychology}
  \bibinfo{volume}{131}, \bibinfo{pages}{367--372}.
\bibitem[{Harrell(1991b)}]{harrell1991precautionary}
\bibinfo{author}{Harrell, W.A.}, \bibinfo{year}{1991}b.
\newblock \bibinfo{title}{Precautionary street crossing by elderly
  pedestrians}.
\newblock \bibinfo{journal}{The International Journal of Aging and Human
  Development} \bibinfo{volume}{32}, \bibinfo{pages}{65--80}.
\bibitem[{Holland and Hill(2007)}]{holland2007effect}
\bibinfo{author}{Holland, C.}, \bibinfo{author}{Hill, R.},
  \bibinfo{year}{2007}.
\newblock \bibinfo{title}{The effect of age, gender and driver status on
  pedestrians’ intentions to cross the road in risky situations}.
\newblock \bibinfo{journal}{Accident Analysis \& Prevention}
  \bibinfo{volume}{39}, \bibinfo{pages}{224--237}.
\bibitem[{Hulse et~al.(2018)Hulse, Xie and Galea}]{hulse2018perceptions}
\bibinfo{author}{Hulse, L.M.}, \bibinfo{author}{Xie, H.},
  \bibinfo{author}{Galea, E.R.}, \bibinfo{year}{2018}.
\newblock \bibinfo{title}{Perceptions of autonomous vehicles: Relationships
  with road users, risk, gender and age}.
\newblock \bibinfo{journal}{Safety Science} \bibinfo{volume}{102},
  \bibinfo{pages}{1--13}.
\bibitem[{Ishaque and Noland(2008)}]{ishaque2008behavioural}
\bibinfo{author}{Ishaque, M.M.}, \bibinfo{author}{Noland, R.B.},
  \bibinfo{year}{2008}.
\newblock \bibinfo{title}{Behavioural issues in pedestrian speed choice and
  street crossing behaviour: a review}.
\newblock \bibinfo{journal}{Transport Reviews} \bibinfo{volume}{28},
  \bibinfo{pages}{61--85}.
\bibitem[{Jayaraman et~al.(2019)Jayaraman, Creech, Dawn, Yang, Pradhan, Tsui,
  Robert et~al.}]{jayaraman2019pedestrian}
\bibinfo{author}{Jayaraman, S.}, \bibinfo{author}{Creech, C.},
  \bibinfo{author}{Dawn, T.}, \bibinfo{author}{Yang, X.J.},
  \bibinfo{author}{Pradhan, A.}, \bibinfo{author}{Tsui, K.},
  \bibinfo{author}{Robert, L.}, et~al., \bibinfo{year}{2019}.
\newblock \bibinfo{title}{Pedestrian trust in automated vehicles: Role of
  traffic signal and av driving behavior} .
\bibitem[{Jennett et~al.(2008)Jennett, Cox, Cairns, Dhoparee, Epps, Tijs and
  Walton}]{jennett2008measuring}
\bibinfo{author}{Jennett, C.}, \bibinfo{author}{Cox, A.L.},
  \bibinfo{author}{Cairns, P.}, \bibinfo{author}{Dhoparee, S.},
  \bibinfo{author}{Epps, A.}, \bibinfo{author}{Tijs, T.},
  \bibinfo{author}{Walton, A.}, \bibinfo{year}{2008}.
\newblock \bibinfo{title}{Measuring and defining the experience of immersion in
  games}.
\newblock \bibinfo{journal}{International journal of human-computer studies}
  \bibinfo{volume}{66}, \bibinfo{pages}{641--661}.
\bibitem[{Kadali et~al.(2014)Kadali, Rathi and Perumal}]{kadali2014evaluation}
\bibinfo{author}{Kadali, B.R.}, \bibinfo{author}{Rathi, N.},
  \bibinfo{author}{Perumal, V.}, \bibinfo{year}{2014}.
\newblock \bibinfo{title}{Evaluation of pedestrian mid-block road crossing
  behavior using an artificial neural network (ann)}, in:
  \bibinfo{booktitle}{CICTP 2014: Safe, Smart, and Sustainable Multimodal
  Transportation Systems}, pp. \bibinfo{pages}{1911--1922}.
\bibitem[{Kalantarov et~al.(2018)Kalantarov, Riemer and
  Oron-Gilad}]{kalantarov2018pedestrians}
\bibinfo{author}{Kalantarov, S.}, \bibinfo{author}{Riemer, R.},
  \bibinfo{author}{Oron-Gilad, T.}, \bibinfo{year}{2018}.
\newblock \bibinfo{title}{Pedestrians’ road crossing decisions and body
  parts’ movements}.
\newblock \bibinfo{journal}{Transportation research part F: traffic psychology
  and behaviour} \bibinfo{volume}{53}, \bibinfo{pages}{155--171}.
\bibitem[{Kalatian and Farooq(2019)}]{kalatian2019deepwait}
\bibinfo{author}{Kalatian, A.}, \bibinfo{author}{Farooq, B.},
  \bibinfo{year}{2019}.
\newblock \bibinfo{title}{Deepwait: Pedestrian wait time estimation in mixed
  traffic conditions using deep survival analysis}, in:
  \bibinfo{booktitle}{2019 IEEE Intelligent Transportation Systems Conference
  (ITSC)}, \bibinfo{organization}{IEEE}. pp. \bibinfo{pages}{2034--2039}.
\bibitem[{Kalatian et~al.(2018)Kalatian, Sobhani and Farooq}]{kalatian2018cox}
\bibinfo{author}{Kalatian, A.}, \bibinfo{author}{Sobhani, A.},
  \bibinfo{author}{Farooq, B.}, \bibinfo{year}{2018}.
\newblock \bibinfo{title}{Analysis of distracted pedestrians’ waiting time:
  Head-mounted immersive virtual reality application}, in:
  \bibinfo{booktitle}{Proceedings of Pedestrian and Evacuation Dynamics 2018},
  \bibinfo{address}{Lund, Sweden}.
\bibitem[{Kaplan and Meier(1958)}]{kaplan1958nonparametric}
\bibinfo{author}{Kaplan, E.L.}, \bibinfo{author}{Meier, P.},
  \bibinfo{year}{1958}.
\newblock \bibinfo{title}{Nonparametric estimation from incomplete
  observations}.
\newblock \bibinfo{journal}{Journal of the American statistical association}
  \bibinfo{volume}{53}, \bibinfo{pages}{457--481}.
\bibitem[{Katzman et~al.(2016)Katzman, Shaham, Cloninger, Bates, Jiang and
  Kluger}]{katzman2016deep}
\bibinfo{author}{Katzman, J.L.}, \bibinfo{author}{Shaham, U.},
  \bibinfo{author}{Cloninger, A.}, \bibinfo{author}{Bates, J.},
  \bibinfo{author}{Jiang, T.}, \bibinfo{author}{Kluger, Y.},
  \bibinfo{year}{2016}.
\newblock \bibinfo{title}{Deep survival: A deep cox proportional hazards
  network}.
\newblock \bibinfo{journal}{stat} \bibinfo{volume}{1050}, \bibinfo{pages}{2}.
\bibitem[{Kirkpatrick et~al.(1983)Kirkpatrick, Gelatt and
  Vecchi}]{kirkpatrick1983optimization}
\bibinfo{author}{Kirkpatrick, S.}, \bibinfo{author}{Gelatt, C.D.},
  \bibinfo{author}{Vecchi, M.P.}, \bibinfo{year}{1983}.
\newblock \bibinfo{title}{Optimization by simulated annealing}.
\newblock \bibinfo{journal}{science} \bibinfo{volume}{220},
  \bibinfo{pages}{671--680}.
\bibitem[{Kononenko et~al.(1997)Kononenko, {\v{S}}imec and
  Robnik-{\v{S}}ikonja}]{kononenko1997overcoming}
\bibinfo{author}{Kononenko, I.}, \bibinfo{author}{{\v{S}}imec, E.},
  \bibinfo{author}{Robnik-{\v{S}}ikonja, M.}, \bibinfo{year}{1997}.
\newblock \bibinfo{title}{Overcoming the myopia of inductive learning
  algorithms with relieff}.
\newblock \bibinfo{journal}{Applied Intelligence} \bibinfo{volume}{7},
  \bibinfo{pages}{39--55}.
\bibitem[{Konstantinou et~al.(2014)Konstantinou, Biedermann and
  Kimber}]{konstantinou2014optimal}
\bibinfo{author}{Konstantinou, M.}, \bibinfo{author}{Biedermann, S.},
  \bibinfo{author}{Kimber, A.}, \bibinfo{year}{2014}.
\newblock \bibinfo{title}{Optimal designs for two-parameter nonlinear models
  with application to survival models}.
\newblock \bibinfo{journal}{Statistica Sinica} \bibinfo{volume}{24},
  \bibinfo{pages}{415--428}.
\bibitem[{Lipovetsky and Conklin(2001)}]{lipovetsky2001analysis}
\bibinfo{author}{Lipovetsky, S.}, \bibinfo{author}{Conklin, M.},
  \bibinfo{year}{2001}.
\newblock \bibinfo{title}{Analysis of regression in game theory approach}.
\newblock \bibinfo{journal}{Applied Stochastic Models in Business and Industry}
  \bibinfo{volume}{17}, \bibinfo{pages}{319--330}.
\bibitem[{Lipton(2016)}]{lipton2016mythos}
\bibinfo{author}{Lipton, Z.C.}, \bibinfo{year}{2016}.
\newblock \bibinfo{title}{The mythos of model interpretability}.
\newblock \bibinfo{journal}{arXiv preprint arXiv:1606.03490} .
\bibitem[{Luck et~al.(2017)Luck, Sylvain, Cardinal, Lodi and
  Bengio}]{luck2017deep}
\bibinfo{author}{Luck, M.}, \bibinfo{author}{Sylvain, T.},
  \bibinfo{author}{Cardinal, H.}, \bibinfo{author}{Lodi, A.},
  \bibinfo{author}{Bengio, Y.}, \bibinfo{year}{2017}.
\newblock \bibinfo{title}{Deep learning for patient-specific kidney graft
  survival analysis}.
\newblock \bibinfo{journal}{arXiv preprint arXiv:1705.10245} .
\bibitem[{Lundberg and Lee(2017)}]{lundberg2017unified}
\bibinfo{author}{Lundberg, S.M.}, \bibinfo{author}{Lee, S.I.},
  \bibinfo{year}{2017}.
\newblock \bibinfo{title}{A unified approach to interpreting model
  predictions}, in: \bibinfo{booktitle}{Advances in Neural Information
  Processing Systems}, pp. \bibinfo{pages}{4765--4774}.
\bibitem[{Mahadevan et~al.(2018)Mahadevan, Somanath and
  Sharlin}]{mahadevan2018communicating}
\bibinfo{author}{Mahadevan, K.}, \bibinfo{author}{Somanath, S.},
  \bibinfo{author}{Sharlin, E.}, \bibinfo{year}{2018}.
\newblock \bibinfo{title}{Communicating awareness and intent in autonomous
  vehicle-pedestrian interaction}, in: \bibinfo{booktitle}{Proceedings of the
  2018 CHI Conference on Human Factors in Computing Systems}, pp.
  \bibinfo{pages}{1--12}.
\bibitem[{Mariani et~al.(1997)Mariani, Coradini, Biganzoli, Boracchi, Marubini,
  Pilotti, Salvadori, Silvestrini, Veronesi, Zucali
  et~al.}]{mariani1997prognostic}
\bibinfo{author}{Mariani, L.}, \bibinfo{author}{Coradini, D.},
  \bibinfo{author}{Biganzoli, E.}, \bibinfo{author}{Boracchi, P.},
  \bibinfo{author}{Marubini, E.}, \bibinfo{author}{Pilotti, S.},
  \bibinfo{author}{Salvadori, B.}, \bibinfo{author}{Silvestrini, R.},
  \bibinfo{author}{Veronesi, U.}, \bibinfo{author}{Zucali, R.}, et~al.,
  \bibinfo{year}{1997}.
\newblock \bibinfo{title}{Prognostic factors for metachronous contralateral
  breast cancer: a comparison of the linear cox regression model and its
  artificial neural network extension}.
\newblock \bibinfo{journal}{Breast cancer research and treatment}
  \bibinfo{volume}{44}, \bibinfo{pages}{167--178}.
\bibitem[{Millard-Ball(2018)}]{millard2018pedestrians}
\bibinfo{author}{Millard-Ball, A.}, \bibinfo{year}{2018}.
\newblock \bibinfo{title}{Pedestrians, autonomous vehicles, and cities}.
\newblock \bibinfo{journal}{Journal of Planning Education and Research}
  \bibinfo{volume}{38}, \bibinfo{pages}{6--12}.
\bibitem[{Molnar(2019)}]{molnar2019interpretable}
\bibinfo{author}{Molnar, C.}, \bibinfo{year}{2019}.
\newblock \bibinfo{title}{Interpretable machine learning}.
\newblock \bibinfo{publisher}{Lulu. com}.
\bibitem[{Nah et~al.(2011)Nah, Eschenbrenner and DeWester}]{nah2011enhancing}
\bibinfo{author}{Nah, F.F.H.}, \bibinfo{author}{Eschenbrenner, B.},
  \bibinfo{author}{DeWester, D.}, \bibinfo{year}{2011}.
\newblock \bibinfo{title}{Enhancing brand equity through flow and telepresence:
  A comparison of 2d and 3d virtual worlds}.
\newblock \bibinfo{journal}{MIs Quarterly} , \bibinfo{pages}{731--747}.
\bibitem[{Ortuzar and Willumsen(1994)}]{ortuzar1994modelling}
\bibinfo{author}{Ortuzar, J.d.}, \bibinfo{author}{Willumsen, L.G.},
  \bibinfo{year}{1994}.
\newblock \bibinfo{title}{Modelling transport}.
\bibitem[{Oxley et~al.(2005)Oxley, Ihsen, Fildes, Charlton and
  Day}]{oxley2005crossing}
\bibinfo{author}{Oxley, J.A.}, \bibinfo{author}{Ihsen, E.},
  \bibinfo{author}{Fildes, B.N.}, \bibinfo{author}{Charlton, J.L.},
  \bibinfo{author}{Day, R.H.}, \bibinfo{year}{2005}.
\newblock \bibinfo{title}{Crossing roads safely: an experimental study of age
  differences in gap selection by pedestrians}.
\newblock \bibinfo{journal}{Accident Analysis \& Prevention}
  \bibinfo{volume}{37}, \bibinfo{pages}{962--971}.
\bibitem[{Pillai et~al.(2017)}]{pillai2017virtual}
\bibinfo{author}{Pillai, A.}, et~al., \bibinfo{year}{2017}.
\newblock \bibinfo{title}{Virtual reality based study to analyse pedestrian
  attitude towards autonomous vehicles} .
\bibitem[{Rasouli et~al.(2017)Rasouli, Kotseruba and
  Tsotsos}]{rasouli2017agreeing}
\bibinfo{author}{Rasouli, A.}, \bibinfo{author}{Kotseruba, I.},
  \bibinfo{author}{Tsotsos, J.K.}, \bibinfo{year}{2017}.
\newblock \bibinfo{title}{Agreeing to cross: How drivers and pedestrians
  communicate}, in: \bibinfo{booktitle}{2017 IEEE Intelligent Vehicles
  Symposium (IV)}, \bibinfo{organization}{IEEE}. pp. \bibinfo{pages}{264--269}.
\bibitem[{Rasouli and Tsotsos(2019)}]{rasouli2019autonomous}
\bibinfo{author}{Rasouli, A.}, \bibinfo{author}{Tsotsos, J.K.},
  \bibinfo{year}{2019}.
\newblock \bibinfo{title}{Autonomous vehicles that interact with pedestrians: A
  survey of theory and practice}.
\newblock \bibinfo{journal}{IEEE transactions on intelligent transportation
  systems} .
\bibitem[{Ribeiro et~al.(2016)Ribeiro, Singh and Guestrin}]{ribeiro2016should}
\bibinfo{author}{Ribeiro, M.T.}, \bibinfo{author}{Singh, S.},
  \bibinfo{author}{Guestrin, C.}, \bibinfo{year}{2016}.
\newblock \bibinfo{title}{Why should i trust you?: Explaining the predictions
  of any classifier}, in: \bibinfo{booktitle}{Proceedings of the 22nd ACM
  SIGKDD international conference on knowledge discovery and data mining},
  \bibinfo{organization}{ACM}. pp. \bibinfo{pages}{1135--1144}.
\bibitem[{Robnik-{\v{S}}ikonja and Kononenko(1997)}]{robnik1997adaptation}
\bibinfo{author}{Robnik-{\v{S}}ikonja, M.}, \bibinfo{author}{Kononenko, I.},
  \bibinfo{year}{1997}.
\newblock \bibinfo{title}{An adaptation of relief for attribute estimation in
  regression}, in: \bibinfo{booktitle}{Machine Learning: Proceedings of the
  Fourteenth International Conference (ICML’97)}, pp.
  \bibinfo{pages}{296--304}.
\bibitem[{Robnik-{\v{S}}ikonja and Kononenko(2003)}]{robnik2003theoretical}
\bibinfo{author}{Robnik-{\v{S}}ikonja, M.}, \bibinfo{author}{Kononenko, I.},
  \bibinfo{year}{2003}.
\newblock \bibinfo{title}{Theoretical and empirical analysis of relieff and
  rrelieff}.
\newblock \bibinfo{journal}{Machine learning} \bibinfo{volume}{53},
  \bibinfo{pages}{23--69}.
\bibitem[{Rose and Bliemer(2009)}]{rose2009constructing}
\bibinfo{author}{Rose, J.M.}, \bibinfo{author}{Bliemer, M.C.},
  \bibinfo{year}{2009}.
\newblock \bibinfo{title}{Constructing efficient stated choice experimental
  designs}.
\newblock \bibinfo{journal}{Transport Reviews} \bibinfo{volume}{29},
  \bibinfo{pages}{587--617}.
\bibitem[{Rose et~al.(2008)Rose, Bliemer, Hensher and
  Collins}]{rose2008designing}
\bibinfo{author}{Rose, J.M.}, \bibinfo{author}{Bliemer, M.C.},
  \bibinfo{author}{Hensher, D.A.}, \bibinfo{author}{Collins, A.T.},
  \bibinfo{year}{2008}.
\newblock \bibinfo{title}{Designing efficient stated choice experiments in the
  presence of reference alternatives}.
\newblock \bibinfo{journal}{Transportation Research Part B: Methodological}
  \bibinfo{volume}{42}, \bibinfo{pages}{395--406}.
\bibitem[{Schmidt and Schwabe(2015)}]{schmidt2015optimal}
\bibinfo{author}{Schmidt, D.}, \bibinfo{author}{Schwabe, R.},
  \bibinfo{year}{2015}.
\newblock \bibinfo{title}{On optimal designs for censored data}.
\newblock \bibinfo{journal}{Metrika} \bibinfo{volume}{78},
  \bibinfo{pages}{237--257}.
\bibitem[{Schmidt and Faerber(2009)}]{schmidt2009pedestrians}
\bibinfo{author}{Schmidt, S.}, \bibinfo{author}{Faerber, B.},
  \bibinfo{year}{2009}.
\newblock \bibinfo{title}{Pedestrians at the kerb--recognising the action
  intentions of humans}.
\newblock \bibinfo{journal}{Transportation research part F: traffic psychology
  and behaviour} \bibinfo{volume}{12}, \bibinfo{pages}{300--310}.
\bibitem[{Street et~al.(2005)Street, Burgess and Louviere}]{street2005quick}
\bibinfo{author}{Street, D.J.}, \bibinfo{author}{Burgess, L.},
  \bibinfo{author}{Louviere, J.J.}, \bibinfo{year}{2005}.
\newblock \bibinfo{title}{Quick and easy choice sets: constructing optimal and
  nearly optimal stated choice experiments}.
\newblock \bibinfo{journal}{International Journal of Research in Marketing}
  \bibinfo{volume}{22}, \bibinfo{pages}{459--470}.
\bibitem[{{\v{S}}trumbelj and Kononenko(2014)}]{vstrumbelj2014explaining}
\bibinfo{author}{{\v{S}}trumbelj, E.}, \bibinfo{author}{Kononenko, I.},
  \bibinfo{year}{2014}.
\newblock \bibinfo{title}{Explaining prediction models and individual
  predictions with feature contributions}.
\newblock \bibinfo{journal}{Knowledge and information systems}
  \bibinfo{volume}{41}, \bibinfo{pages}{647--665}.
\bibitem[{Sun et~al.(2003)Sun, Ukkusuri, Benekohal and
  Waller}]{sun2003modeling}
\bibinfo{author}{Sun, D.}, \bibinfo{author}{Ukkusuri, S.},
  \bibinfo{author}{Benekohal, R.F.}, \bibinfo{author}{Waller, S.T.},
  \bibinfo{year}{2003}.
\newblock \bibinfo{title}{Modeling of motorist-pedestrian interaction at
  uncontrolled mid-block crosswalks}, in: \bibinfo{booktitle}{Transportation
  Research Record, TRB Annual Meeting CD-ROM, Washington, DC}.
\bibitem[{Sun et~al.(2015)Sun, Zhuang, Wu, Zhao and Zhang}]{sun2015estimation}
\bibinfo{author}{Sun, R.}, \bibinfo{author}{Zhuang, X.}, \bibinfo{author}{Wu,
  C.}, \bibinfo{author}{Zhao, G.}, \bibinfo{author}{Zhang, K.},
  \bibinfo{year}{2015}.
\newblock \bibinfo{title}{The estimation of vehicle speed and stopping distance
  by pedestrians crossing streets in a naturalistic traffic environment}.
\newblock \bibinfo{journal}{Transportation research part F: traffic psychology
  and behaviour} \bibinfo{volume}{30}, \bibinfo{pages}{97--106}.
\bibitem[{Therneau and Grambsch(2013)}]{therneau2013modeling}
\bibinfo{author}{Therneau, T.M.}, \bibinfo{author}{Grambsch, P.M.},
  \bibinfo{year}{2013}.
\newblock \bibinfo{title}{Modeling survival data: extending the Cox model}.
\newblock \bibinfo{publisher}{Springer Science \& Business Media}.
\bibitem[{Tom and Grani{\'e}(2011)}]{tom2011gender}
\bibinfo{author}{Tom, A.}, \bibinfo{author}{Grani{\'e}, M.A.},
  \bibinfo{year}{2011}.
\newblock \bibinfo{title}{Gender differences in pedestrian rule compliance and
  visual search at signalized and unsignalized crossroads}.
\newblock \bibinfo{journal}{Accident Analysis \& Prevention}
  \bibinfo{volume}{43}, \bibinfo{pages}{1794--1801}.
\bibitem[{Wang et~al.(2018)Wang, Wang and Zhao}]{wang2018deep}
\bibinfo{author}{Wang, S.}, \bibinfo{author}{Wang, Q.}, \bibinfo{author}{Zhao,
  J.}, \bibinfo{year}{2018}.
\newblock \bibinfo{title}{Deep neural networks for choice analysis: Extracting
  complete economic information for interpretation}.
\newblock \bibinfo{journal}{arXiv preprint arXiv:1812.04528} .
\bibitem[{Wang et~al.(2011)Wang, Guo, Gao and Bubb}]{wang2011individual}
\bibinfo{author}{Wang, W.}, \bibinfo{author}{Guo, H.}, \bibinfo{author}{Gao,
  Z.}, \bibinfo{author}{Bubb, H.}, \bibinfo{year}{2011}.
\newblock \bibinfo{title}{Individual differences of pedestrian behaviour in
  midblock crosswalk and intersection}.
\newblock \bibinfo{journal}{International Journal of Crashworthiness}
  \bibinfo{volume}{16}, \bibinfo{pages}{1--9}.
\bibitem[{Wong and Farooq(2020)}]{wong2020bi}
\bibinfo{author}{Wong, M.}, \bibinfo{author}{Farooq, B.}, \bibinfo{year}{2020}.
\newblock \bibinfo{title}{A bi-partite generative model framework for analyzing
  and simulating large scale multiple discrete-continuous travel behaviour
  data}.
\newblock \bibinfo{journal}{Transportation Research Part C: Emerging
  Technologies} \bibinfo{volume}{110}, \bibinfo{pages}{247--268}.
\bibitem[{Yagil(2000)}]{yagil2000beliefs}
\bibinfo{author}{Yagil, D.}, \bibinfo{year}{2000}.
\newblock \bibinfo{title}{Beliefs, motives and situational factors related to
  pedestrians’ self-reported behavior at signal-controlled crossings}.
\newblock \bibinfo{journal}{Transportation Research Part F: Traffic Psychology
  and Behaviour} \bibinfo{volume}{3}, \bibinfo{pages}{1--13}.
\bibitem[{Yu et~al.(2011)Yu, Greiner, Lin and Baracos}]{yu2011learning}
\bibinfo{author}{Yu, C.N.}, \bibinfo{author}{Greiner, R.},
  \bibinfo{author}{Lin, H.C.}, \bibinfo{author}{Baracos, V.},
  \bibinfo{year}{2011}.
\newblock \bibinfo{title}{Learning patient-specific cancer survival
  distributions as a sequence of dependent regressors}, in:
  \bibinfo{booktitle}{Advances in Neural Information Processing Systems}, pp.
  \bibinfo{pages}{1845--1853}.

\end{thebibliography}

\appendix
\section{Braking levels for automated vehicles}
\label{A:brake}
\arash{Three levels of brake for the automated vehicles are defined in our VR experiment as follows:
\begin{itemize}
    \item \textbf{Level 1}: given the initial speed of the vehicle when observing the pedestrian ($V_0$) and its corresponding distance to the pedestrian ($d$), \textit{one} constant deceleration rate is calculated to stop the car before the pedestrian and avoid a collision. A maximum possible deceleration rate is set to be $a_{max} = -3 m/s^2$. Thus, if the calculated acceleration rate surpasses the maximum possible rate, the vehicle stops with the rate of $a_{max}$, but cannot stop for the pedestrian completely in a timely manner. 
    \item \textbf{Level 2:} given the initial speed of the vehicle when observing the pedestrian ($V_0$) and its corresponding distance to the pedestrian ($d$), \textit{two} constant deceleration rates are calculated $a_1$ and $a_2$. $a_1$ changes the speed of vehicle from $V_0$ to $V_0/2$ in $d/2$ of the distance, and $a_2$ decreases the speed from there to stop in the rest $d/2$ of distance. Similar to level 1, deceleration rate cannot exceed $a_{max}$ in neither of the two steps. 
    \item \textbf{Level 3:} given the initial speed of the vehicle when observing the pedestrian ($V_0$) and its corresponding distance to the pedestrian ($d$), \textit{three} constant deceleration rates are calculated: $a_1$, $a_2$, and $a_3$. $a_1$ changes the speed of vehicle from $V_0$ to $2V_0/3$ in the first $d/3$ of the distance, $a_2$ decreases the speed from there to $V_0/3$ in the next $d/3$ of the distance, and $a_3$ stops the vehicle in the final $d/3$ of distance. Similar to level 1 and 3, deceleration rate cannot exceed $a_{max}$ in none of the steps. 
\end{itemize}}
\arash{Based on the definitions provided, the speed-distance profile of the braking levels is drawn for three scenarios:}
\begin{itemize}
    \item \textbf{Scenario 1:} $V_0 = 40 km/hr$, $d = 40 meters$ 
    
     \item \textbf{Scenario 2:} $V_0 = 50 km/hr$, $d = 40 meters$ 
     
     \item \textbf{Scenario 3:} $V_0 = 50 km/hr$, $d = 20 meters$ 
\end{itemize}

\begin{figure}[!t]
\begin{adjustbox}{varwidth=\textwidth,fbox,center}
\centering
\begin{subfigure}{0.95\linewidth}

    \begin{subfigure}[b]{0.45\textwidth}
         \centering
         \includegraphics[scale=0.5]{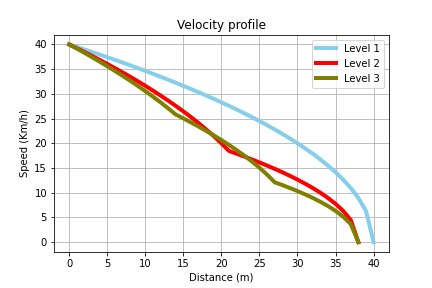}
         \caption{Initial speed: 40 km/hr, initial distance: 40 m}
     \end{subfigure}
     \hfill
     \begin{subfigure}[b]{0.45\textwidth}
         \centering
         \includegraphics[scale=0.5]{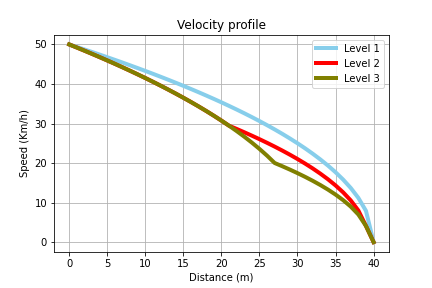}
         \caption{Initial speed: 50 km/hr, initial distance: 40 m}
     \end{subfigure}
\end{subfigure}\\[2ex]
\begin{subfigure}{0.8\linewidth}

\centering
\includegraphics[scale=0.5]{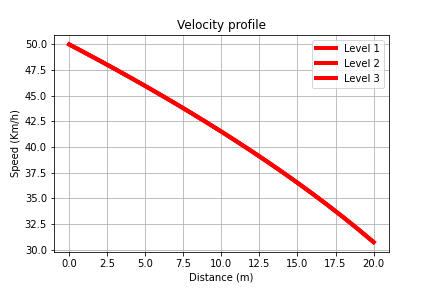}
\caption{Initial speed: 50 km/hr, initial distance: 20 m}

\end{subfigure}
\end{adjustbox}
\caption{Velocity profile for three hypothetical scenarios}
\label{fig:prof}
\end{figure}

\arash{In scenario 1, the vehicle with level 1 braking system manages to decelerate and stop the car right before a barrier seen in 40 meters, whereas the two other braking levels stop the vehicle a few meters before the barrier. As the initial speed increases to 50 km/hr in scenario 2, level 2 and level 3 also take longer distance to stop the car, although the speed of the vehicle is slower when it is close to the barrier, generating a more smooth experience for the pedestrian. In scenario 3, however, not enough distance exists between the barrier and the vehicle. The 20 meter distance makes all the braking levels at all their steps to perform at $a_{max}$, making all similar profiles for all the levels. The speed profile of the three aforementioned scenarios are presented in Figure~\ref{fig:prof}.}

\section{Description of covariates}
\label{A:var}
A description of all the variables that were collected during our experiments through VIRE and questionnaire, as well as the possible values they can take are presented in Table~\ref{tab:vars}.
\begin{table}
\caption{Description on the collected covariates}
\small
\begin{tabular}{|ccc|}
\hline
\textbf{Variable}            & \textbf{Description}                                                                                                                                       & \textbf{Range}                                                                            \\ \hline\hline
\multicolumn{3}{|l|}{\textit{Collected through VR}}                                                                                                                                                                                                                                   \\ \hline\hline
Speed Limit (km/hr)          & \begin{tabular}[c]{@{}l@{}}Maximum speed allowed for the vehicles\\ in the traffic simulation\end{tabular}                                                 & 30, 40, 50 km/hr                                                                          \\ \hline
Density                      & Number of vehicles per km in traffic simulation                                                                                                            & 10 to 35 veh/km                                                                           \\ \hline
Minimum allowed gap time     & \begin{tabular}[c]{@{}l@{}}Minimum gap time allowed between the vehicles\\ in traffic simulation\end{tabular}                                              & 1, 1.5, 2 s                                                                               \\ \hline
Lane width                   & Width of one lane that the pedestrian is facing                                                                                                            & 2.5, 2.75, 3 m                                                                            \\ \hline
Road type                    & \begin{tabular}[c]{@{}l@{}}Whether the cross is at a one-way one-lane road,  two-way, \\ two-lane road or two-way two-lane road with a median\end{tabular} & Categorical                                                                               \\ \hline
No. of barking levels        & \begin{tabular}[c]{@{}l@{}}For AVs: levels in which the vehicle decelerates , \\ higher values mean   smoother braking\end{tabular}                        & Categorical                                                                               \\ \hline
Traffic automation status    & \begin{tabular}[c]{@{}l@{}}If all the vehicles in traffic are human-driven,\\ fully automated or a mixture of both\end{tabular}                            & Categorical                                                                               \\ \hline
Arrival rate                 & Number of cars generated in the traffic simulation per hour                                                                                                & 530, 750, 1100 veh/hr                                                                     \\ \hline
Time of day                  & If the scenarios are in the daylight or at night                                                                                                           & Day/ Night                                                                                \\ \hline
Weather                      & \begin{tabular}[c]{@{}l@{}}If the scenarios are in clear weather \\ condition or in snowy environments\end{tabular}                                        & Clear/Snowy                                                                               \\ \hline\hline
\multicolumn{3}{|l|}{\textit{Collected through questionnaire}}                                                                                                                                                                                                                        \\ \hline\hline
Age                          & Bracket of age for the participant,                                                                                                                        & \begin{tabular}[c]{@{}l@{}}{[}18-29{]}, {[}30, 39{]},\\ {[}40.49{]}, over 50\end{tabular} \\ \hline
Gender                       & Gender of the participant                                                                                                                                  & Male, Female                                                                              \\ \hline
Occupation                   & Occupation of the participant                                                                                                                              & Categorical                                                                               \\ \hline
Education                    & Highest level of education for the participant                                                                                                             & Categorical                                                                               \\ \hline
Walt to work                 & Whether the participant usually walks for commute or not                                                                                                   & Binary                                                                                    \\ \hline
Walk for shopping            & \begin{tabular}[c]{@{}l@{}}Whether the participant usually walks for\\  buying grocery or not\end{tabular}                                                 & Binary                                                                                    \\ \hline
Driving license              & Whether the participant owns a driving license                                                                                                             & Binary                                                                                    \\ \hline
No. of cars in the household & Number of cars the participant have in their household                                                                                                     & Categorical (0,1, more than 1)                                                            \\ \hline
Main mode of transportation  & Active modes: bike and walk, Public transit, Private Car                                                                                                   & Categorical                                                                               \\ \hline
Previous VR experience       & \begin{tabular}[c]{@{}l@{}}If the participant has used VR in\\ any form previously at least one time\end{tabular}                                          & Binary                                                                                    \\ \hline
\end{tabular}
\label{tab:vars}
\end{table}
\section{D-optimal design for Cox Proportional Hazards (CPH) model}
\label{A:design}
\cite{schmidt2015optimal} proposed the D-Optimal design for a CPH with three covariates. In this section, we extend and generalize their formulation for cases where the number of covariates in the Hazards model can be more than 3. Cox proportional hazards model with fixed termination time $c$, is specified as: 
\begin{linenomath}
\begin{equation}
    T_k \sim \exp\Big(\exp( \beta Z_k)\Big)
\end{equation}
\end{linenomath}
where $T_k$ is the survival time for instance (participant) $k$, \(Z_k = \begin{bmatrix}
1 & Z_{1k} & ... &Z_{ik}
\end{bmatrix} \) is the vector of covariate values of instance k with first element added for the intercept, and $\mathbf{\beta}$ is the vector of parameters associated with $i$ independent variables. If the data are type I censored, meaning that the experiment stops at a predetermined time which is the case of our experiment, the Fisher information matrix is given by~\citep{konstantinou2014optimal}:
\begin{linenomath}
\begin{equation}
\mathbf{I}(Z_k,\beta)=\Big(1- \exp (-c \exp(\beta Z_k ))\Big)Z_k Z_k^T
\end{equation} 
\end{linenomath}
Suppose that we need to generate $m$  scenarios for the experiment, then
\(
\xi = \begin{bmatrix}
\mathbf{S_1} & \mathbf{S_2} & ... & \mathbf{S_m}\\
\omega_1 & \omega_2 &... & \omega_m
\end{bmatrix}
\) is the design of experiment where
$\mathbf{S_m}$ represents the $m^{th}$ combination of all the independent variables and $(0\leq \omega_m \leq 1)$ is the associated weight, where $\sum_m\omega_m = 1$. For such model the information matrix will be:
\begin{linenomath}
\begin{equation}
 \mathbf{M}(\xi, \mathbf{\beta}) = \sum_{j=1}^m \omega_j \mathbf{I}(\mathbf{S_j},\beta)
\end{equation}
\end{linenomath}
Given $\beta$, $Z$, and $m$, we need to find a design $\xi_{\beta}^*$ that maximizes $\det (\mathbf{M}(\xi, \mathbf{\beta}))$. For a high number of independent variables, we can use Markov Chain Monte Carlo simulation for finding $\xi_{\beta}^*$. In particular, Simulated Annealing (SA) is a great candidate for such problems \citep{kirkpatrick1983optimization}. Using prior $\beta$ coefficients collected in our trial data collection, having all the possible levels of covariates, and setting the number of experiments to 90, we run a simulated annealing to maximize the determinant of Fisher information matrix of our design. Algorithm~\ref{sa} provides the pseudo-code for the simulated annealing we used in this study.
\begin{algorithm}
\SetKwData{Left}{left}\SetKwData{This}{this}\SetKwData{Up}{up}
\SetKwFunction{Union}{Union}\SetKwFunction{FindCompress}{FindCompress}
\SetKwInOut{Input}{input}\SetKwInOut{Output}{output}
Set $\beta := \beta_0$ \\
Set total number of covariates $N_v$\\
Set possible values that covariates can take $P$\\
Set total number of experiments $N_e$ \\
Select an initial random design $\xi = \xi_0$ and  weights $W = W_0$ based on $N_v$, $P$ \& $N_e$. \\
Calculate $\det (\mathbf{M}(\xi_0, \mathbf{\beta}))$ and set $\xi_0$ and its weights as the best design ($\xi^*$,  $W^*$)\\
Set maximum number of iterations $n$, initial temperature $T_0$ and temperature reduction rate $\alpha$\\
$T = T_0$\\
\For{$i\leftarrow 1$ \KwTo $n$}{
\emph{select a neighbour design $\xi_n$ or neighbour weights ${W_n}$}\;
\emph{\uIf {$\det (\mathbf{M}(\xi_{n}, \mathbf{\beta})) > \det(\mathbf{M}(\xi^{*}, \mathbf{\beta}))$} {$\xi^* = \xi_n \: \& \: W^*=W_n$}
\Else{ $\Delta = \displaystyle \frac{(\det (\mathbf{M}(\xi^{*}, \mathbf{\beta})) - \det(\mathbf{M}(\xi_{n}, \mathbf{\beta})))}{\det (\mathbf{M}(\xi^{*},\mathbf{\beta})} $\\
$H=exp(-\Delta/T)$\\
\uIf{H $\geq $\text{U}(0,1)}{$\xi^{*} = \xi_n \: \& \: W^* = W_n$}}
}
\emph{$T = \alpha \times T$ }
}
Return $(\xi^{*}$, $W^*)$
\caption{Pseudo-code for the simulated annealing algorithm}\label{sa}
\end{algorithm}\DecMargin{1em}
\section{Descriptive statistics}
\label{A:data}

\begin{small}
\begin{longtable}{|ccccc|}
\caption{Wait time for different levels of covariates}\\
\hline
\textbf{Covariate}                        & \textbf{Level}  & \textbf{Avg wait time   (s)} & \textbf{std wait time   (s)} & \textbf{p-value} \\ \hline\hline
\endhead
\multirow{3}{*}{Automation  Status}       & FAV             & 3.9987                       & 4.4798                       & Base             \\ 
                                          & HDV             & 3.7836                       & 3.8286                       & 0.722            \\ 
                                          & Mixed           & 4.5966                       & 4.5626                       & 0.217            \\ \hline
\multirow{3}{*}{Speed Limit}              & 30              & 4.3901                       & 4.8337                       & Base             \\ 
                                          & 40              & 4.0629                       & 4.3745                       & 0.163            \\ 
                                          & 50              & 3.5984                       & 4.1429                       & 0.001            \\ \hline
\multirow{3}{*}{Lane Width}               & 2.5             & 3.5040                       & 4.2706                       & Base             \\ 
                                          & 2.75            & 4.1080                       & 4.4896                       & 0.011            \\ 
                                          & 3               & 4.4562                       & 4.5940                       & 0.000            \\ \hline
\multirow{3}{*}{Gap Time}                 & 1               & 3.9540                       & 4.3206                       & Base             \\ 
                                          & 1.5             & 3.9859                       & 4.4149                       & 0.896            \\ 
                                          & \textit{2}      & \textit{4.0993}              & 4.6364                       & 0.532            \\ \hline
\multirow{3}{*}{Arrival Rate}             & 530             & 3.5911                       & 4.0619                       & Base             \\ 
                                          & 750             & 4.0488                       & 4.4978                       & 0.045            \\ 
                                          & 1100            & 4.4928                       & 4.8377                       & 0.000            \\ \hline
\multirow{3}{*}{Density}                  & High            & 4.6726                       & 4.9279                       & Base             \\ 
                                          & Low             & 3.4576                       & 3.8202                       & 0.000            \\ 
                                          & Medium          & 4.0013                       & 4.5642                       & 0.005            \\ \hline
\multirow{3}{*}{Braking Levels}           & 1               & 3.7320                       & 4.3655                       & base             \\ 
                                          & 2               & 4.1022                       & 4.4905                       & 0.126            \\ 
                                          & 3               & 4.1729                       & 4.5810                       & 0.072            \\ \hline
\multirow{3}{*}{Road Type}                & One way         & 4.3061                       & 4.5648                       & Base             \\ 
                                          & Two way         & 4.0539                       & 4.6069                       & 0.281            \\ 
                                          & With Median     & 3.6817                       & 4.1925                       & 0.008            \\ \hline
\multirow{3}{*}{Time}                     & Day             & 3.9869                       & 4.4601                       & Base             \\ 
                                          & Night           & 4.0770                       & 4.4840                       & 0.657            \\ 
                                          & HDV             & 4.2826                       & 4.2986                       & 0.178            \\ \hline
\multirow{2}{*}{Weather}                  & Clear           & 3.9399                       & 4.3597                       & 0.000            \\ 
                                          & Snowy           & 4.1903                       & 4.6960                       & 0.229            \\ \hline
\multirow{4}{*}{Age}                      & 18-29           & 4.4550                       & 4.5867                       & Base             \\ 
                                          & 30-39           & 3.4088                       & 4.1414                       & 0.000            \\ 
                                          & 40-49           & 3.4420                       & 4.2984                       & 0.006            \\ 
                                          & Over 50         & 4.8674                       & 4.9639                       & 0.260            \\ \hline
\multirow{2}{*}{Gender}                   & Female          & 4.1904                       & 4.6353                       & Base             \\ 
                                          & Male            & 3.8917                       & 4.3382                       & 0.125            \\ \hline
\multirow{2}{*}{Driving License}          & No              & 4.9646                       & 4.7475                       & Base             \\ 
                                          & Yes             & 3.9330                       & 4.4329                       & 0.003            \\ \hline
\multirow{3}{*}{\#Cars Owned}             & 0               & 3.6999                       & 4.1485                       & Base             \\ 
                                          & 1               & 3.9531                       & 4.2400                       & 0.327            \\ 
                                          & \textgreater{}1 & 4.2464                       & 4.8043                       & 0.030            \\ \hline
\multirow{4}{*}{Main Mode}                & Bike            & 4.2048                       & 4.6735                       & Base             \\ 
                                          & Car             & 4.1805                       & 4.6745                       & 0.938            \\ 
                                          & Public          & 3.7670                       & 4.1270                       & 0.143            \\ 
                                          & Walking         & 4.3526                       & 4.9069                       & 0.716            \\ \hline
\multirow{3}{*}{Walk to Work}             & No              & 3.8725                       & 4.4066                       & Base             \\ 
                                          & Sometimes       & 4.0767                       & 4.5276                       & 0.388            \\ 
                                          & Yes             & 4.4434                       & 4.5787                       & 0.030            \\ \hline
\multirow{3}{*}{Walk to Shop}             & No              & 4.3275                       & 4.7820                       & Base             \\ 
                                          & Sometimes       & 4.2457                       & 4.5664                       & 0.787            \\ 
                                          & Yes             & 3.8040                       & 4.2712                       & 0.020            \\ \hline
\multirow{2}{*}{Previous VR   Experience} & No              & 4.0596                       & 4.4724                       & Base             \\ 
                                          & Yes             & 3.9612                       & 4.4626                       & 0.613            \\ \hline
\end{longtable}
\end{small}

\end{document}